\documentclass[preprint,showpacs,preprintnumbers,amsmath,amssymb,superscriptaddress,nofootinbib]{revtex4}
\usepackage{graphicx,color}
\usepackage{amsmath,amssymb}
\usepackage{url}
\usepackage{epstopdf}
\newcommand{\hs}{\hspace*{0.5cm}}

\newcommand{\be}{\begin{equation}}
\newcommand{\ee}{\end{equation}}
\newcommand{\bea}{\begin{eqnarray}}
\newcommand{\eea}{\end{eqnarray}}
\newcommand{\nn}{\nonumber}
\newcommand{\crn}{\nonumber \\}
\newcommand{\non}{\nonumber}

\newcommand{\al}{\alpha}
\newcommand{\la}{\lambda}
\newcommand{\bet}{\beta}
\newcommand{\ga}{\gamma}

\newcommand{\fr}{\frac}
\newcommand{\fb}{\fbox}
\newcommand{\bc}{\begin{center}}
\newcommand{\ec}{\end{center}}

\newcommand {\ba}{\begin{array}}
\newcommand {\ea}{\end{array}}
\newcommand{\ben}{\begin{enumerate}}
\newcommand{\een}{\end{enumerate}}

\usepackage{epsfig,graphicx}
\usepackage{bm}
\usepackage{dcolumn}

\begin{document}

\title{Lepton flavor violating decays of Standard-Model-like Higgs in 3-3-1 model with neutral lepton}

\author{L. T. Hue}\email{lthue@iop.vast.ac.vn}
\affiliation{Institute of Physics,   Vietnam Academy of Science and Technology, 10 Dao Tan, Ba
Dinh, Hanoi, Vietnam }

\author{H. N. Long}\email{hoangngoclong@tdt.edu.vn}
\affiliation{Theoretical Particle Physics and Cosmology Research Group, Ton Duc Thang University, Ho Chi Minh City, Vietnam; 
Faculty of Applied Sciences,
Ton Duc Thang University, Ho Chi Minh City, Vietnam.}

\author{T. T. Thuc}\email{ttthuc@grad.iop.vast.ac.vn}
\affiliation{Institute of Physics, VAST, 10 Dao Tan, Ba Dinh, Hanoi, Vietnam}

\author{T. Phong Nguyen}\email{ thanhphong@ctu.edu.vn}
\affiliation{Department of Physics, Cantho University,
 3/2 Street, Ninh Kieu, Cantho, Vietnam}

\begin{abstract}
The one loop contribution to the lepton flavor violating decay $h^0\rightarrow \mu\tau$ of the SM-like neutral Higgs (LFVHD) in the 3-3-1 model with neutral lepton is calculated using the unitary gauge. We have checked in detail that the total contribution is exactly finite, and the divergent cancellations happen separately in two  parts of active neutrinos and exotic heavy leptons. By numerical investigation, we have indicated that the one-loop contribution of the active neutrinos is very suppressed while that of exotic leptons is rather large.  The branching ratio of the LFVHD  strongly depends  on the Yukawa couplings between exotic leptons and $SU(3)_L$ Higgs triplets.  This ratio can reach $10^{-5}$ providing large Yukawa couplings and constructive correlations of the $SU(3)_L$ scale ($v_3$) and the charged Higgs masses.  The branching ratio decreases rapidly with the small  Yukawa couplings and large $v_3$.
\end{abstract}
\pacs{12.15.-y, 14.60.-z,  14.80.Ec
 }
\maketitle


\section{Introduction}
\label{sec:intro}
The observation  the Higgs boson with mass around 125.09  GeV  by experiments at  the Large Hadron
Collider (LHC) \cite{higgsdicovery1,higgsdicovery2,h2ga1,h2ga2,comHiggs} again confirms the very success of the Standard Model (SM) at low energies of below few hundred GeV. But the SM must be extended to solve many well-known problems,  at least the question  of  neutrino masses and neutrino oscillations which have been experimentally confirmed \cite{nnpro}. Neutrino oscillation is a  clear evidence of lepton flavor violation in the neutral lepton sector which may give loop contributions to the rare lepton flavor violating (LFV) decays of charged leptons,  $Z$  and SM-like Higgs bosons.  Therefore, these are  the promoting  subjects of new physics which have been hunted by recent experiments \cite{exLFVl,exLFVz,exLFVh}.  Especially, the latest experimental results of LFVHD  have been reported  recently by CMS and ATLAS. Defining Br$(h^0\rightarrow\mu\tau)\equiv\mathrm{Br}(h^0\rightarrow\mu^+\tau^-) +\mathrm{Br}(h^0\rightarrow\mu^-\tau^+)$, the upper bound  Br$(h^0\rightarrow\mu\tau) < 1.5\times 10^{-2}$ at 95\% C.L. was announced by  CMS, in agreement with $1.85\times 10^{-2}$ at 95\% C.L. from ATLAS. These sensitivities are not far from the recent theoretical prediction and is hoped to be improved soon,  as discussed in \cite{MainHll}.

  The LFVHD of  the  neutral Higgses  have been investigated widely in the well-known models beyond the SM \cite{LFVgeneral,FHgauge,MainHll}, including the  supersymmetric (SUSY) models \cite{Anna1,LFVHSUSY331,massInsert1}.  The SUSY versions usually  predict large branching ratio of LFVHD which can reach $10^{-4}$ or higher, even  up to $10^{-2}$ in recent investigation \cite{Anna1}, provided the two following requirements:  new LFV sources from sleptons and  the large $\tan\beta$-ratio of two vacuum expectation values (vev) of two neutral Higgses. At least it is true for the LFVHD $h^0\rightarrow \mu\tau$ under the restrict of the recent upper bound of  Br$(\tau\rightarrow\mu\gamma)<10^{-8}$\cite{extmga2}.  In   the non-SUSY $SU(2)_L\times U(1)_Y$ models beyond the SM such as the seesaw or general two Higgs doublet (THDM),  the LFVHD still depends on the LFV decay of $\tau$ lepton. The reason is that  the LFVHD is strongly affected by Yukawa couplings of leptons while the $SU(2)_L\times U(1)_Y$ contains only small Yukawa couplings of normal charged leptons and active neutrinos. Therefore, many of non-SUSY versions  predict the suppressed signal of LFVHD.

Based on the extension of the $SU(2)_L\times U(1)_Y$ gauge symmetry of the SM to the $SU(3)_L\times U(1)_X$,   there is a class of models called 3-3-1 models which  inherit new LFV sources.  Firstly,  the particle spectra include new charged gauge bosons and charged Higgses, normally carrying two units of
lepton number.   Secondly, the third components of the lepton (anti-) triplets may be   normal  charged leptons  \cite{pisanom331,framton} or  new leptons   \cite{prtests331,newHeavyL,newLihgtL,331custodial,snu331} with non-zero lepton numbers. These new leptons can mix among one to another to create new LFV changing currents, except the case of normal charged leptons. The most interesting models for LFVHD are the ones  with new heavy leptons corresponding to new Yukawa couplings that affect strongly to the LFVHD through the loop contributions. This property is different from the models based on the gauge symmetry of the SM including the SUSY versions. In the 3-3-1 models, if the new particles and  the $SU(3)_L$ scale are larger than few hundred GeVs, the one-loop contributions to  the LFV decays of $\tau$ always satisfy the recent experimental bound \cite{lfv331}.  While this region of parameter space,  even at the TeV values of the $SU(3)_L$ scale,  favors the large branching ratios of LFVHD. The one-loop contributions on LFV processes in SUSY versions of 3-3-1 models were given in \cite{tLFV331,LFVHSUSY331}, but the non-SUSY contributions were not mentioned.

 The 3-3-1 models were first investigated from interest of the simplest expansion of the $SU(2)_L$ gauge symmetry and the simplest lepton sector \cite{pisanom331}.  They then became more attractive by a clue of answering the flavor question coming  from the requirement of anomaly cancellation for $SU(3)_L\times U(1)_X$ gauge symmetry \cite{framton}. The violation of the lepton number is a natural property of these models, leading to the natural presence of the  LFV processes and neutrino oscillations. Many versions of 3-3-1 models have been constructed for explaining other unsolved questions in the SM limit: solving the strong CP problem \cite{CP331} with Peccei-Quinn symmetry \cite{PQ331}; allowing the electric charge quantization \cite{echarge331},...  More interesting, the neutral heavy leptons or neutral Higgses can play roles of candidates of dark matter (DM) \cite{snu331}.  Besides, the models with neutral leptons are still  interesting  for investigation of precision tests \cite{prtests331}.

From the above reasons, this work will pay attention to the LFVHD  of the 3-3-1 with left-handed heavy neutral leptons or neutrinos (3-3-1LHN) \cite{snu331}.  It is then easy to predict which specific 3-3-1 models can give large signals of LFVHD.  As we will see, the 3-3-1 models usually contain new heavy neutral Higgses, including both CP-even and odd ones. But the recent lower bound of the $SU(3)_L$ scale is  few TeV, resulting the same order of   these Higgs masses.  At recent collision energies of experiments, the opportunity to observe  these heavy neutral Higgses seems rare. We therefore concentrate  only on  the SM-like Higgs.

Our work is arranged as follows. The section \ref{ffactor} will pay attention on the formula of branching ratio of LFVHD which  can be also applied for new neutral CP-even Higgses, listing the Feynman rules and the  needed form factors to calculate the amplitudes for general 3-3-1 models.  In the section \ref{lfv331n},  the model constructed in \cite{snu331} will be improved including adding new LFV couplings;  imposing a custodial symmetry on the Higgs potential to cancel large flavor neutral changing currents in the Higgs sector and  simplify the Higgs self-interactions.  From this  both masses and mass eigenvectors of even-CP neutral Higgses are found exactly at the tree level.  The section \ref{num} represents numerical results of LFVHD, where  the most interesting region of the parameter space will be chosen based on the latest experimental results relating to lower bounds of new gauge bosons and charged Higgses. We concentrate on the roles of  Yukawa couplings of exotic neutral leptons, the charged Higgses and the $SU(3)_L$ scale.  We summarize our main results in the conclusion section.  The appendices show  notations of Passarino-Veltman functions, the detail  of calculating one-loop contributions to  LFVHD amplitude in the 3-3-1LHN and the  divergent  cancellation.

 \section{\label{ffactor}Formulas for decay rates of neutral Higgses}
\begin{figure}[h]
  \includegraphics*[width=13cm]{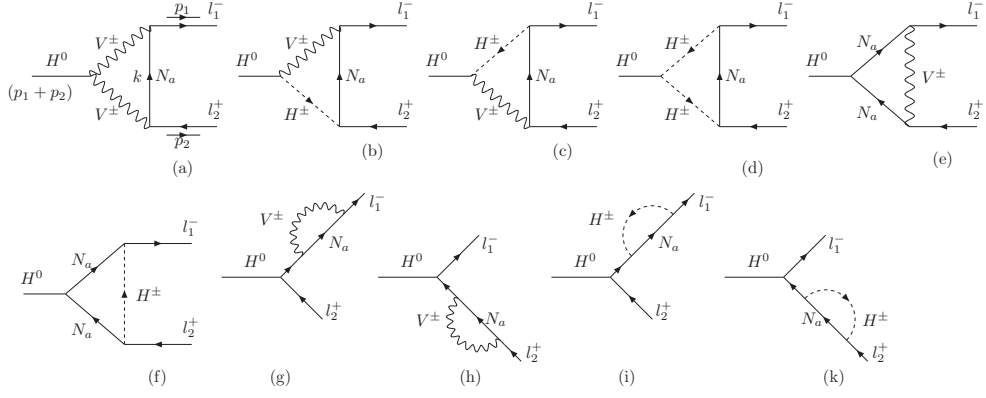}
  \caption{Feynman diagrams contributing to the $H^0\rightarrow \mu^{\pm}\tau^{\mp}$ decay  in the unitary gauge, where $H^0$ is an arbitrary even-CP neutral Higgs in the 3-3-1 models, including the SM-like one.}\label{nDiagram}
\end{figure}

 For studying the LFVHD, namely $h^0\rightarrow \tau^{\pm}\mu^{\mp}$, we consider the general form of the  corresponding LFV effective Lagrangian as follows
 \be  - \mathcal{L}^{LFV}= h^0\left(\Delta_L \overline{\mu}P_L \tau +\Delta_R \overline{\mu}P_R \tau\right) + \mathrm{h.c.},\label{Lfvh0eff}\ee
  where   $\Delta_{L,R}$ are scalar factors arisen from the loop contributions. In the unitary gauge,  the one-loop diagrams contributing to $\Delta_{L,R}$  are listed in the figure \ref{nDiagram}. They can be applied for  the models beyond the SM where the particle contents include only Higgses, fermions  and gauge bosons.
  The amplitude  decay  is \cite{MainHll}:
\be
i\mathcal{M}=-i\bar{u}_{1}\left(\Delta_LP_L+\Delta_RP_R\right)v_{2},
\ee
where $u_1\equiv u_1(p_1,s_1)$ and $v_2\equiv v_2(p_2,s_2)$ are respective Dirac spinors of the $\mu$ and $\tau$.
The partial width of  the decays is
{\small \bea
\Gamma (h^0\rightarrow \mu\tau)&\equiv&\Gamma (h^0\rightarrow \mu^{-} \tau^{+})+\Gamma (h^0\rightarrow \mu^{+} \tau^{-})\crn
&= & \fr{1}{8\pi m_{h^0}}\times \sqrt{\left[1-\left(\fr{m_{1}+m_{2}}{m_{h^0}}\right)^2 \right]\left[1-\left(\fr{m_{1}-m_{2}}{m_{h^0}}\right)^2 \right]}\crn
&\times& \left[\fr{}{}\left(m^2_{h^0}-m^2_{1}-m^2_{2}\right) \left(\vert \Delta_L\vert^2+\vert \Delta_R\vert^2\right)-
4m_1m_2\mathrm{Re} \left(\Delta_L\Delta^*_R \right)\right], \label{LFVwidth}
\eea}
where $m_{h^0}$, $m_1$ and $m_2$ are the masses of the neutral Higgs $h^0$, muon and tauon, respectively. They satisfy the on-shell conditions for external particles,
 namely $p^2_{i}=m_i^2$ (i=1,2) and $ p_0^2 \equiv( p_1+p_2)^2=m^2_{h^0}$.

\begin{figure}[h]
  \includegraphics*[width=14cm]{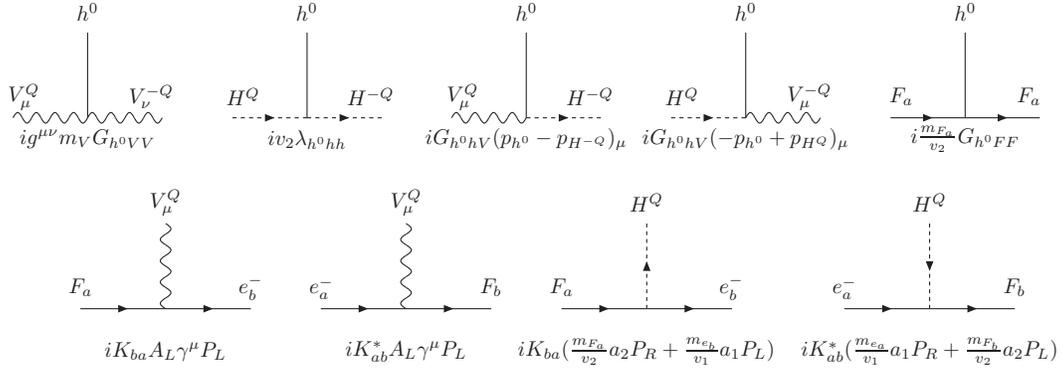}
  \caption{Feynman rules for the $h^0\rightarrow\mu^{\pm}\tau^{\mp}$ in the unitary gauge, where all momenta are incoming}\label{FeynRule}
\end{figure}

In the unitary gauge, the relevant Feynman rules for the LFV decay of $h^0\rightarrow l^{\pm}_1l^{\mp}_2$ are represented in the figure \ref{FeynRule}.

For each diagram, there is a corresponding generic function expressing its contribution to the LFVHD.  These functions are defined as
{\small\bea  E^{FVV}_L(m_F,m_V)
&=&m_V m_1\left\{ \frac{1}{2m_V^4}\left[m_F^2(b^{(1)}_1-b^{(1)}_0-b^{(2)}_0)\right.\right.\crn
 \hs &-&\left.\left. m_2^2b^{(2)}_1 + \left(2m_V^2+m^2_{h^0}\right)m_F^2\left(C_0-C_1\right)\right]\right.\crn
&&\left.-\left(2+\frac{m_1^2-m_2^2}{m_V^2}\right) C_1 +
 \left(\frac{m_1^2-m^2_{h^0}}{m_V^2}+ \frac{m_2^2 m^2_{h^0}}{2m_V^4}\right)C_2\right\}, \label{EfvvL} \\
 E^{FVV}_R(m_F,m_V)&=&m_V m_2\left\{\frac{1}{2 m_V^4}\left[-m_F^2\left(b^{(2)}_1+ b^{(1)}_0 + b^{(2)}_0 \right) \right.\right.\crn
  &+& \left.\left.  m_1 ^2 b^{(1)}_1  +   (2m_V^2+m^2_{h^0}) m_F^2(C_0+C_2)\right] \right.\crn
&&\left.+\left(2+\frac{-m_1^2+m_2^2}{m_V^2}\right)C_2-\left( \frac{m_2^2-m^2_{h^0}}{m_V^2}+ \frac{m_1^2 m^2_{h^0}}{m_V^4}\right)C_1\right\},  \label{EfvvR}
\eea
\bea
 && E^{FVH}_L(a_1,a_2,v_1,v_2,m_F,m_V,m_H)\crn&=&
 m_1\left\{-\fr{a_2}{v_2} \fr{m_F^2}{m_V^2}\left(b^{(1)}_1-b^{(1)}_0\right)   + \fr{a_1}{v_1}m_2^2\left[2 C_1-\left(1+ \fr{m^2_{h}-m^2_{h^0}}{m_V^2}\right) C_2\right]\right.\crn
&&\left.+\fr{a_2}{v_2}m_F^2\left[C_0+C_1+\fr{m^2_{h}-m^2_{h^0}}{m_V^2}\left(C_0-C_1\right) \right]\right\}, \label{EfvhL} \\
&& E^{FVH}_Ra_1,a_2,v_1,v_2,m_F,m_V,m_H)\crn &=&m_2\left\{\fr{a_1}{v_1}\left[\fr{m_1^2b^{(1)}_1-m_F^2b^{(1)}_0}{m_V^2} +\left(\frac{}{}m_F^2C_0-m_1^2C_1+2 m_2^2C_2\right.\right.\right.\crn
 &&\left.\left.+2(m^2_{h^0}-m_2^2)C_1-  \fr{m^2_{h}-m^2_{h^0}}{m_V^2}\left(m^2_FC_0-m_1^2C_1\right)\right)\right]\crn &&+\left.\fr{a_2}{v_2} m_F^2\left(-2C_0-C_2+\fr{m^2_{h}-m^2_{h^0}}{m_V^2}C_2 \right) \right\},   \label{EfvhR}
 \eea
 \bea
  && E^{FHV}_L(a_1,a_2,v_1,v_2,m_F,m_H,m_V)\crn&=& m_1\left\{\fr{a_1}{v_1}\left[\fr{-m_2^2b^{(2)}_1-m_F^2b^{(2)}_0}{m_V^2} +\left(\frac{}{}m_F^2C_0-2m_1^2C_1+ m_2^2C_2\right.\right.\right.\crn
 &&\left.\left.-2(m^2_{h^0}-m_1^2)C_2-  \fr{m^2_{h}-m^2_{h^0}}{m_V^2}\left(m^2_FC_0+m_2^2C_2\right)\right)\right] \crn &&\left.+\fr{a_2}{v_2} m_F^2\left(-2C_0+C_1-\fr{m^2_{h}-m^2_{h^0}}{m_V^2}C_1 \right)\right\}, \label{EfvhL} \\
 && E^{FHV}_R(a_1,a_2,v_1,v_2,m_F,m_H,m_V)\crn&=& m_2 \left\{\fr{a_2}{v_2} \fr{m_F^2}{m_V^2}\left(b^{(2)}_1+b^{(2)}_0\right)
 + \fr{a_1}{v_1}m_1^2\left[-2 C_2+\left(1+ \fr{m^2_{h}-m^2_{h^0}}{m_V^2}\right) C_1\right]\right.\crn
&&\left.+\fr{a_2}{v_2}m_F^2\left[C_0-C_2+\fr{m^2_{h}-m^2_{h^0}}{m_V^2}\left(C_0+C_2\right) \right]\right\}.   \label{EfvhR}
 \eea
 \bea
 E^{FHH}_L(a_1,a_2,v_1,v_2,m_F,m_H)&=&  m_1v_2\left[ \fr{a_1a_2}{v_1v_2}m_F^2C_0-\fr{a^2_1}{v^2_1}m_2^2C_2+\fr{a^2_2}{v^2_2}m_F^2C_1\right] , \crn \label{EfhhL} \\
 E^{FHH}_R(a_1,a_2,v_1,v_2)&=& m_2v_2\left[ \fr{a_1a_2}{v_1v_2}m_F^2C_0+\fr{a^2_1}{v^2_1}m_1^2C_1-\fr{a^2_2}{v^2_2}m_F^2C_2 \right],\crn \label{EfhhR} \eea
 \bea
 E^{VFF}_L(m_V,m_F)&=&\frac{m_1m^2_F}{m_V}\crn
  &\times&\left[\fr{1}{m_V^2}\left(b^{(12)}_{0}
 +b^{(1)}_1 -(m_1^2+m_2^2-2m_F^2)C_1\right)-C_0+4C_1\right],\crn  \label{EvffL} \\
 E^{VFF}_R(m_V,m_F)&=&
 \frac{m_2 m^2_F}{m_V} \crn
 &\times&\left[ \fr{1}{m_V^2}\left(b^{(12)}_{0} -b^{(2)}_1 +(m_1^2+m_2^2-2m_F^2)C_2\right)-C_0-4C_2 \right],\crn \label{EvffR}\eea
\bea
 E^{HFF}_L(a_1,a_2,v_1,v_2)&=&\frac{ m_1m^2_F }{v_2}\crn
 &\times& \left[\dfrac{a_1a_2}{v_1v_2}b^{(12)}_{0}
 +\fr{a_1^2}{v_1^2}m_2^2(2C_2+C_0)+\fr{a_2^2}{v_2^2}m_F^2(C_0-2C_1) \right.\crn
&+& \left.\fr{a_1a_2}{v_1v_2} \left(\frac{}{}2m_2^2C_2-(m_1^2+m_2^2)C_1+(m_F^2+m^2_{h}+m_2^2)C_0\right)\right],\crn  \label{EhffL} \\
 E^{HFF}_R(a_1,a_2,v_1,v_2)&=& \frac{m_2 m^2_F}{v_2}\crn
  &\times&\left[ \dfrac{a_1a_2}{v_1v_2}b^{(12)}_{0}+ \dfrac{a_1^2}{v_1^2}m_1^2(C_0-2C_1)+\fr{a_2^2}{v_2^2}m_F^2(C_0+2C_2)\right.\crn
&+&\left. \fr{a_1a_2}{v_1v_2}\left(\frac{}{}-2m_1^2C_1+(m_1^2+m_2^2)C_2+(m_F^2+m^2_{h}+m_1^2)C_0 \right)\right], \crn\label{EhffR} \eea
 \bea
 E^{FV}_L(m_F,m_V)&=& \fr{-m_1m_2^2}{m_V(m_1^2-m_2^2)}\left[\left(2+\frac{m_F^2}{m_V^2}\right) \left(b^{(1)}_1 +b^{(2)}_1 \right) \right. \crn&+&\left.\fr{m_1^2 b^{(1)}_1 +m_2^2 b^{(2)}_1}{m_V^2} - \fr{2m_F^2}{m_V^2}\left(b^{(1)}_0-b^{(2)}_0\right)\right],  \label{DfvL} \\
 E^{FV}_R(m_F,m_V)&=& \frac{m_1}{m_2}E^{FV}_L, \label{DfvR}\eea
\bea
  E^{FH}_L(a_1,a_2,v_1,v_2)&=& \fr{m_1}{v_1(m_1^2-m_2^2)}\left[m^2_2 \left(m^2_1\fr{a_1^2}{v_1^2}+m^2_F\fr{a^2_2}{v^2_2} \right)
\left(b_1^{(1)}+b_1^{(2)}\right) \right.\crn
&&\left.  \hspace{1.8 cm}+m^2_F\fr{a_1a_2}{v_1v_2}\left(2m^2_2b_0^{(1)}-(m^2_1 +m^2_2)b_0^{(2)}\right) \right], \label{DfhL} \\
 E^{FH}_R(a_1,a_2,v_1,v_2)&=&  \fr{m_2}{v_1(m_1^2-m_2^2)}\left[ m^2_1 \left(m^2_2\fr{a_1^2}{v_1^2}+m^2_F\fr{a^2_2}{v^2_2} \right)\left(b_1^{(1)}+b_1^{(2)}\right)\right.\crn
&&\left.  \hspace{1.8 cm}+m^2_F\fr{a_1a_2}{v_1v_2}\left(-2m^2_1b_0^{(2)}+(m^2_1 +m^2_2)b_0^{(1)}\right)\right]. \label{DfhR} \eea}
The notations are introduced as follows.  All the $b-$  and $C-$ functions are defined in the appendix \ref{mainte}, where  $C-$ functions are well-known as Passarino-Veltman (PV) functions of one-loop three points and $b$-functions are the finite parts of the two-point functions. For convenience,  $m_{e_a}$ and $m_{e_b}$ in the Feynman rules are denoted as $m_1, m_2$,   corresponding to the masses of the final leptons in the LFV decays $h^0\rightarrow l_1^-l_2^+$.   Other parameters are masses of the neutral Higgs $m_{h^0}$,  and the virtual particles in the loops, including gauge boson mass $m_V$, charged Higgs mass $m_{h}$ and fermion masses $m_F$. Specially, the masses of the virtual fermions are denoted as $m_a\equiv m_F$ for convenience. The parameters $a_1,a_2,v_1$ and $v_2$ given in the Feynman rules in the figure \ref{FeynRule}, where $v_1,v_2$ are VEVs giving masses for normal and exotic leptons/active neutrinos; $a_1,a_2$ relate the mixing parameters of the charged Higgses coupling with these leptons.

The set of the form factors (\ref{EfvvL}-\ref{DfhR}) was calculated in details in the appendix  \ref{Camlitude} which we find them consistent with calculations using Form \cite{form}.  These form factors are simpler than those calculated in the appendix because they  contain only terms contributing to the final amplitude of the LFVHD.  The excluded terms are come from the two reasons: i)  those do not contain the neutral leptons in the loop so they vanish after summing all virtual leptons, reflecting the GIM mechanism; ii) the divergent terms defined by (\ref{divt}). The second is true only when  the final contribution is assumed to be finite. This is right for the models having no tree level LFV couplings of $\mu-\tau$. The 3-3-1  LHN model we will consider in this work satisfies this condition and the divergent cancellation is checked precisely in the appendix  \ref{Camlitude}. Another remark is that the divergent term  (\ref{divt}) contains a conventional choice of $\ln\mu^2/m_h^2$ in which $m_{h}$ can be replaced by an arbitrary fixed scale.  We find that only the contributions of the diagram  \ref{nDiagram}d)  and sum of two diagrams   \ref{nDiagram}g) and \ref{nDiagram}h) are finite.

Now the form factors $\Delta_{L,R}$ can be written as  the sum of all $E_{L,R}$ functions. The one loop contributions  to the LFV decays such as $\Delta_{L,R}$ are finite  without using any renormalization procedure  to cancel divergences.  In addition, $\Delta_{L,R}$ do not depend on the  $\mu$ parameter  arising from the dimensional regularization method used to derive  all above scalar $E_{L,R}$ functions in this work. But in general  contributions from the separate  diagrams in the figure \ref{nDiagram} do contain the divergences and therefore the particular finite parts  $E_{L,R}$ do depend on $\mu$, so it will be nonsense for computing separate  contributions.

 Using  the Feynman- 't Hooft gauge, similar expressions of the LFVHD amplitudes as functions of PV-function were introduced in \cite{FHgauge,MainHll}. They were applied for LFVHD in the seesaw models,  where there are no new contributions from new physical charged Higgses or new gauge bosons. The contributions in this case correspond to those of  only four diagrams a),  e) g) and h) in the figure  \ref{nDiagram} of this work. So choosing the unitary gauge is more advantageous for calculating LFVHD predicted by models having complicated particle spectra.

 There is another simple analytic expressions given details in \cite{massInsert1}, updated from previous works \cite{massInsert2}.  It can be applied for not only SUSY models but also the models predicting new  heavy scales including 3-3-1 models. The point is that this treatment uses the $C$-functions with approximation of zero-external momentums of the two charged leptons, i.e. $p_1^2=p_2^2=0$. Unlike the case of LFV decays of $\tau\rightarrow\mu\gamma$, the LFVHD contains a large external momentum of neutral Higgs: $2p_1.p_2\simeq|(p_1\pm p_2)^2|=m_h^2\sim \mathcal{O}([100 \mathrm{GeV}]^2)$, which should be included in the $C$-functions, as discussed in the appendix \ref{mainte}. This is consistent with  discussion on $C$-functions given in \cite{bardin}.

 \section{\label{lfv331n} 3-3-1 model with new neutral lepton}
  In this section we will review a particular 3-3-1 model used to investigate the LFVHD,  namely the 3-3-1LHN \cite{snu331}.  We will keep most of all ingredients shown in ref. \cite{snu331}, while add two new assumptions: i) in order to appear the LFV effects, we assume that apart from the oscillation of the active neutrinos, there also exists  the maximal mixing  in the new lepton sector;  ii) The Higgs potential satisfies a custodial symmetry shown in \cite{331custodial} to avoid large loop contributions of the Higgses to precision tests such as $\rho$-parameter and flavor neutral changing currents.  More interesting, the latter results a very simple Higgs potential in the sense that many independent  Higgs self-couplings are reduced and the squared mass matrix of the neutral Higgses can be solved exactly at the tree level. The following will review the needed ingredients for calculating the LFV decay of $h^0\rightarrow  l_i^+l_j^-$.
 \subsection{Particle content}
 \begin{itemize}
   \item  Fermion.  In each family, all left-handed leptons  are included in the  $SU(3)_L$ triplets while right-handed ones are always singlets,
 \bea L'_{a}=\left(
              \begin{array}{c}
                \nu'_a \\
                e'_a \\
                N'_a \\
              \end{array}
            \right)_L \sim \left(1,3,-\frac{1}{3}\right),\hs e'_{aR}\sim (1,1,-1),\hs N'_{aR}\sim(1,1,0),
  \label{lepton}\eea
  where the numbers in the parentheses are  the respective representations of the $SU(3)_C$, $SU(2)_L$  and $U(1)_X$ gauge groups. The prime denotes the lepton in the flavor basis. Recall that as one of the  assumption in \cite{snu331},  the active neutrinos have no right-handed components and their Majorana masses are generated from the effective dimension-five operators. There is no mixing among active neutrinos and exotic neutral leptons.
   \item Gauge boson. The $SU(3)_L\times U(1)_X$ includes 8 gauge bosons $W^a_{\mu}$ (a=1,8) of the $SU(3)_L$ and the $X_{\mu}$ of the $U(1)_X$, corresponding to eight $SU(3)_L$ generators $T^a$ and a $U(1)_X$ generator $T^9$. The respective covariant derivative is
   \be D_{\mu}\equiv \partial_{\mu}-i g_3 W^a_{\mu}T^a-g_1 T^9X X_{\mu}. \label{code} \ee
    Denote the Gell-Mann matrices as $\lambda_a$, we have $T^a=\frac{1}{2}\lambda_a,-\frac{1}{2}\lambda_a^T$  or $0$ depending on the triplet, antitriplet or singlet representation of the $SU(3)_L$ that  $T^a$ acts on. The $T^9$ is defined as $T^9=\frac{1}{\sqrt{6}}$ and $X$ is the $U(1)_X$ charge of the field it acts on.
   \item Higgs. The model includes three Higgs triplets,
   \bea \rho=\left(
                   \begin{array}{c}
                     \rho_1^+ \\
                     \rho^0 \\
                     \rho_2^+ \\
                   \end{array}
                 \right)\sim \left(1,3,\frac{2}{3}\right), \hs \eta=\left(
                                                       \begin{array}{c}
                                                         \eta_1^0 \\
                                                         \eta^- \\
                                                         \eta_2^0 \\
                                                       \end{array}
                                                     \right),\;
                                                     \chi=\left(
                                                       \begin{array}{c}
                                                         \chi_1^0 \\
                                                         \chi^- \\
                                                         \chi_2^0 \\
                                                       \end{array}
                                                     \right)\sim \left(1,3,-\frac{1}{3}\right).   \label{Higgs1}\eea
  As normal, the 3-3-1 model has two breaking steps: $SU(3)_L\times U(1)_X\overbrace{\rightarrow}^{\langle\chi\rangle}$ $SU(2)_L\times U(1)_Y \overbrace{\rightarrow}^{\langle\rho\rangle,\langle\eta\rangle}U(1)_Q$, leading to the limit $|\langle\chi\rangle|\gg |\langle\rho\rangle|, ~|\langle\eta\rangle|$. The non-zero $U(1)_G$ charged field  $\eta_2^0$ and $\chi_1^0$ have zero vacuum expectation (vev) values: $\langle\eta_2^0\rangle=\langle\chi_1^0\rangle=0$, i.e
  \be \eta^0_2\equiv \frac{S'_2+i A'_2}{\sqrt{2}},\hs \chi^0_1\equiv \frac{S_3+i A_3}{\sqrt{2}}. \label{nhiggs1}\ee
   Others neutral Higgs components can be written as
  {\small \be \rho^0 =\frac{1}{\sqrt{2}}\left(v_1+S_1+iA_1\right),\; \eta^0_1=\frac{1}{\sqrt{2}}\left(v_2+S_2+iA_2\right),\; \chi_2^0=\frac{1}{\sqrt{2}}\left(v_3+S'_3+iA'_3\right).\label{vev1}\ee}
As shown in ref. \cite{331custodial}, after the first breaking step, the corresponding Higgs potential of the 3-3-1 model should keep a custodial symmetry to avoid large FCNCs as well as the large deviation of $\rho$-parameter value obtained  from experiment. This only involves to the $\rho$ and $\eta$  Higgs scalars which generate non-zero vevs in the second breaking step. Applying the Higgs potential satisfying the custodial symmetry given in \cite{THcus}, we obtain a  Higgs potential of the form,
\bea \mathcal{V}&=&\mu^2_{1}\left(\rho^{\dagger}\rho+\eta^{\dagger}\eta\right)+\mu^2_2\chi^{\dagger}\chi+\lambda_1\left[\rho^{\dagger}\rho +\eta^{\dagger}\eta \right]^2+\lambda_2\left(\chi^{\dagger}\chi\right)^2 \crn
&+& \lambda_{12} \left(\rho^{\dagger}\rho+\eta^{\dagger}\eta\right) \left(\chi^{\dagger}\chi\right) - \sqrt{2}f\left( \epsilon_{ijk}\rho^i\eta^i\chi^k+ \mathrm{h.c.}\right), \label{331LHNpo}\eea
where $f$ is assumed to be real. Minimizing this potential leads to $v_1=v_2$ and two additional conditions,
\bea  \mu_1^2+ 2\lambda_1 v_1^2+\frac{1}{2}\lambda_{12}v_3^2&=&f v_3,\crn
\mu_2^2+ \lambda_2 v_3^2+\lambda_{12}v_1^2&=&\frac{f v_1^2}{ v_3}.
\label{mincond}\eea
We stress that if the custodial symmetry is kept in this 3-3-1 model, the model automatically satisfies most of  the conditions assumed  in ref. \cite{snu331} for purpose of  simplifying or reducing independent parameters in the Higgs potential.  For this work, which especially concentrates on the neutral Higgses, the most important consequence is that all of the mass basis of Higgses, including the neutral, can be found exactly without reduction of the number of Higgs multiplets.
 \end{itemize}
 In the following, we just pay attention to those  used directly  in this work, i.e. the  mass spectra of leptons, gauge bosons and Higgses. Other parts have been mentioned in \cite{snu331}.
 \subsection{Mass spectra}
 \subsubsection{Leptons}
  We use the Yukawa terms shown in \cite{snu331} for generating masses of charged leptons, active neutrinos and heavy neutral leptons, namely
  \be -\mathcal{L}^{Y}_{\mathrm{lepton}} = y^{e}_{ab}\overline{L'_{a}}\rho e'_{bR}+ y^{N}_{ab}\overline{L'_{a}}\chi N'_{bR}+ \frac{y^{\nu}_{ab}}{\Lambda} \left(\overline{(L'_{a})^c}\eta^*\right)\left(\eta^{\dagger}L'_b\right) + \mathrm{h.c.}, \label{Ylepton1}\ee
  where the notation $(L')^c_a=( (\nu'_{aL})^c,\;(e'_{aL})^c,\;(N'_{aL})^c\;)^T\equiv( \nu'^c_{aR},\;e'^c_{aR},\;N'^c_{aR}\;)^T$  implies that $\psi^c_R\equiv P_R \psi^c= (\psi_L)^c$ with $\psi$ and $\psi^c \equiv C\overline{\psi}^T$ being the Dirac spinor and its charge conjugation, respectively. The $\Lambda$ is some high energy scale.  Remind that $\psi_L= P_L \psi, \; \psi_R=P_R\psi$ where $P_{R,L}\equiv \frac{1\pm\gamma_5}{2}$ are the right- and left-chiral operators.
  The corresponding mass terms are
  \be -\mathcal{L}^{Y}_{\mathrm{lepton}} = \left[ \frac{y^{e}_{ab}v_1}{\sqrt{2}}\overline{e'_{aL}} e'_{bR}+\frac{ y^{N}_{ab}v_3}{\sqrt{2}}\overline{N_{aL}}  N'_{bR}+ \mathrm{h.c.} \right]+ \frac{y^{\nu}_{ab}v^2_2}{2 \Lambda}\left[ (\overline{\nu'^c_{aR}} \nu'_{bL})+ \mathrm{h.c.}\right]. \label{mterm}\ee
  This means that the active neutrinos are pure Majorana spinors corresponding to the mass matrix  $(M_{\nu})_{ab} \equiv  \frac{y^{\nu}_{ab}v^2_2}{ \Lambda}$. This matrix can be proved to be symmetric \cite{neutrinomass} (chapter 4), therefore the mass eigenstates can be found by a single rotation expressed by a  mixing matrix $U$ that satisfies $ U^{\dagger} M_{\nu}U=\mathrm{diagonal}(m_{\nu_1},\;m_{\nu_2}, \;m_{\nu_3})$, where $m_{\nu_i}$ (i=1,2,3) are mass eigenvalues of the active neutrinos.

  Now we define transformations between the flavor basis  $\{e'_{aL,R},~\nu'_{aL},~N'_{aL,R}\}$ and the mass basis $\{e_{aL,R},~\nu_{aL},~N_{aL,R}\}$:
 \be
e'^-_{aL}= e^-_{aL},  ~~e'^-_{aR}=e^-_{aR}, \;
\nu'_{aL}=U_{ab}\nu_{bL},\; N'_{aL}=V^L_{ab}N_{bL},\quad N'_{aR}=V^R_{ab}N_{bR},
\label{lepmixing}\ee
where $V^L_{ab},~U^L_{ab}$ and  $V^R_{ab}$ are transformations between flavor and mass bases of  leptons. Here unprimed fields denote the mass eigenstates.
Remind that $\nu'^c_{aR}=(\nu'_{aL})^c=U_{ab}\nu^c_{aR}$. The four-spinors representing the active neutrinos are $\nu^c_{a}=\nu_{a}\equiv (\nu_{aL},\; \nu^c_{aR})^T$, resulting the following equalities: $\nu_{aL}=P_L\nu^c_a=P_L\nu_a$ and $\nu^c_{aR}=P_R\nu^c_a=P_R\nu_a$. The upper bounds of recent experiments for the LFV processes in the normal charged  leptons are very suppressed \cite{exLFVl}, therefore  suggest that the two flavor and mass bases  of charged leptons should be the same.

The relations between the mass matrices of leptons in  two flavor and mass bases are
\bea m_{e_a}&=&\frac{v_1}{\sqrt{2}}y^e_{a},\hs y^e_{ab}=y^e_a\delta_{ab},\hs a,b=1,2,3,
 \crn \frac{v_2^2}{\Lambda} U^{\dagger}Y^{\nu} U&=& \mathrm{Diagonal}(m_{\nu_1},~m_{\nu_2},~m_{\nu_3}),\crn
 \frac{v_3}{\sqrt{2}} V^{L\dagger}Y^N V^R&=& \mathrm{Diagonal}(m_{N_1},~m_{N_2},~m_{N_3}),\label{cema1} \eea
 where $Y^{\nu}$ and $Y^N$ are Yukawa matrices defined as $(Y^{\nu})_{ab}=y^{\nu}_{ab}$ and $(Y^{N})_{ab}=y^{N}_{ab}$.

 The  Yukawa interactions between leptons and Higgses can be written according to the lepton mass eigenstates,
 {\small \bea  -\mathcal{L}^{Y}_{\mathrm{lepton}} &=&\frac{m_{e_b}}{v_1}\sqrt{2} \left[\rho^{0}_1 \bar{e}_bP_Re_b+  U^{*}_{ba}\bar{\nu}_a P_Re_b\rho_1^{+} + V^{L*}_{ba}\overline{N}_a P_Re_b\rho_2^{+}+\mathrm{h.c.}   \right]\crn
  &&+\frac{m_{N_a}}{v_3}\sqrt{2} \left[\chi^{0}_2 \bar{N}_aP_RN_a+  V^{L}_{ba}\bar{e}_b P_RN_a\chi^{-}+\mathrm{h.c.}   \right]\crn
  &&+ \frac{m_{\nu_a}}{v_2}\left[S_2\overline{\nu_{a}}P_L\nu_{b}+\fr{1}{\sqrt{2}} \eta^{+}\left(U^{*}_{ba} \overline{\nu_{a}}P_Le_{b}+ U_{ba} \overline{e^c_{b}}P_L\nu_{a}\right)+\mathrm{h.c.} \right], \label{llh}\eea}
where we have used the Marojana property of the active neutrinos: $\nu^c_a=\nu_a$ with $a=1,2,3$.  In addition, using the equality $\overline{e^c_{b}}P_L\nu_{a}=  \overline{\nu_{a}}P_Le_{b}$ for this case the term relating with $\eta^{\pm}$ in the last line of (\ref{llh}) is reduced to $\sqrt{2}\eta^{+} \overline{\nu_{a}}P_Le_{b}$.
\subsubsection{Gauge bosons}

It is simpler to write the charged gauge bosons in the form of $W^aT^a$ with $T^a$  being the gamma matrices, namely
\bea W^a_{\mu} T^a= \frac{1}{\sqrt{2}} \left(
                \begin{array}{ccc}
                  0 & W^{+}_{\mu} & U^0_{\mu} \\
                  W^{-}_{\mu} & 0 & V^{-}_{\mu} \\
                  U^{0*}_{\mu} & V^{+}_{\mu} & 0 \\
                \end{array}
              \right).
\label{gaugeboson1} \eea
The masses of these gauge bosons are:
\be  m_W^2=\frac{g^2v^2}{4},\hs m^2_{U}=m^2_{V} =\frac{g^2}{4}\left(v^2_3+\frac{v^2}{2}\right), \label{gmass}\ee
where we have used the relation $v_1=v_2=\frac{v}{\sqrt{2}}$ and the matching condition   of the $W$ boson mass in 3-3-1 model with that of the SM.

The covariant derivatives  of the leptons contain the lepton-lepton-gauge boson couplings, namely
\bea \mathcal{L}^D_{\mathrm{lepton}} &=& i\overline{L'_a}\gamma^{\mu}D_{\mu}L'_a\crn
&\rightarrow& \frac{g}{\sqrt{2}}\left[ U^*_{ba}\overline{\nu_a}\gamma^{\mu}P_L e_bW^+_{\mu} +U_{ab}\overline{e_b}\gamma^{\mu}P_L\nu_aW^-_{\mu}\right. \crn &+&\left. V^{L*}_{ba}\overline{N_a}\gamma^{\mu}P_L e_bV^+_{\mu} +V^L_{ab}\overline{e_b}\gamma^{\mu}P_LN_aV^-_{\mu}  \right] . \label{cdelepton}\eea
\subsubsection{Higgs bosons}
\begin{itemize}
  \item Singly charged Higgses. There are two Goldstone bosons $G^{\pm}_W$ and $G^{\pm}_V$ of the respective singly charged gauge bosons $W^{\pm}$ and $V^{\pm}$. Two other massive singly charged Higgses have masses
      \be m^2_{H_1}= (1+t^2) fv_3,\hs m^2_{H_2}= 2 f v_3,\label{schiggs}\ee
      where $t\equiv \frac{v_1}{v_3}=\frac{v}{v_3\sqrt{2}}=\tan{\theta}$. Denoting $s_{\theta}\equiv\sin\theta,~c_{\theta}\equiv\cos\theta$, we get some useful relations
      \be m_W=\sqrt{2} m_Vs_{\theta}, \hs v_3= \frac{2m_V}{g}c_{\theta},\hs v_1=v_2=\frac{2m_V}{g}s_{\theta} .\label{theta1}\ee  The relation between two  flavor and mass bases of the singly Higgses are
  \bea \left(
         \begin{array}{c}
           \rho^{\pm}_1 \\
           \eta^{\pm} \\
         \end{array}
       \right)= \frac{1}{\sqrt{2}}\left(
                 \begin{array}{cc}
                   -1 & 1 \\
                   1& 1 \\
                 \end{array}
               \right) \left(
         \begin{array}{c}
         G^{\pm}_W \\
           H_2^{\pm} \\
         \end{array}
       \right), \hs  \left(
         \begin{array}{c}
           \rho^{\pm}_2 \\
           \chi^{\pm} \\
         \end{array}
       \right)=\left(
                 \begin{array}{cc}
                   -s_{\theta} & c_{\theta} \\
                   c_{\theta}& s_{\theta} \\
                 \end{array}
               \right) \left(
         \begin{array}{c}
         G^{\pm}_V \\
           H_1^{\pm} \\
         \end{array}
       \right).
   \label{sHiggse}\eea

  \item CP-odd neutral Higgses.  There are three Goldstone bosons $G_{Z}, G_{Z'}$ and $G'_{U^0}$, and two massive CP-odd neutral Higgses $H_{A_1}$ and $H_{A_2}$ with the values of squared masses are
      \be m^2_{A_1}= m^2_{H_1}=\frac{(1+t^2)}{2}m^2_{H_2},\;  m^2_{A_2}= \frac{(2+t^2)}{2}m^2_{H_2}. \label{oddNmass}\ee
      The relations between the two bases are:
      \bea \left(
             \begin{array}{c}
               A_3 \\
               A'_2 \\
             \end{array}
           \right)= \left(
                      \begin{array}{cc}
                        c_{\theta} & -s_{\theta} \\
                        s_{\theta} & c_{\theta} \\
                      \end{array}
                    \right) \left(
                              \begin{array}{c}
                                G_{3} \\
                                H_{A_2} \\
                              \end{array}
                            \right), \; \left(
                                          \begin{array}{c}
                                            A_{1} \\
                                            A'_{3} \\
                                            A_2 \\
                                          \end{array}
                                        \right)\left(
                                                 \begin{array}{ccc}
                                                   -s_{\theta} & \frac{-c^2_{\theta}}{\sqrt{c^2_{\theta}+1}} &  \frac{c_{\theta}}{\sqrt{c^2_{\theta}+1}} \\
                                                    c_{\theta}&\frac{-s_{\theta}c_{\theta}}{\sqrt{c^2_{\theta}+1}}  & \frac{s_{\theta}}{\sqrt{c^2_{\theta}+1}} \\
                                                  0 & \frac{1}{\sqrt{c^2_{\theta}+1}}& \frac{c_{\theta}}{\sqrt{c^2_{\theta}+1}}\\
                                                 \end{array}
                                               \right)\left(
                                                        \begin{array}{c}
                                                          G_1 \\
                                                          G_2 \\
                                                          H_{A_1} \\
                                                        \end{array}
                                                      \right).\crn
       \label{eoddnA}\eea
  \item CP-even neutral Higgses. Apart from the three exactly massive Higgses shown in the ref. \cite{331custodial}, the model predicts one more Goldstone boson $G_{U}$ and another massive Higgs. The masses and egeinstates of these Higgses are
      \bea  m^2_{h^0_1}&=&\frac{v_3^2}{2}\left[ 4\lambda_1t^2+2\lambda_2 + \frac{t^2 f}{v_3}- \sqrt{\Delta}\right], \crn
       m^2_{h^0_2}&=& \frac{v_3^2}{2}\left[ 4\lambda_1t^2+2\lambda_2 + \frac{t^2 f}{v_3}+ \sqrt{\Delta}\right],\crn
      m^2_{h^0_3}&=& m^2_{H^{\pm}_1}, \;  m^2_{h^0_4}=m^2_{A_2},\label{h0mass}\eea
       where $\Delta=\left( 4 \lambda_1 t^2-2\lambda_2 -\frac{ t^2f}{v_3}\right)^2+8t^2\left( \lambda_{12} -\frac{f}{v_3}\right)^2$.  The  transformations among the flavor  and the mass bases are
       \bea \left(
              \begin{array}{c}
                S'_2 \\
                S_{3} \\
              \end{array}
            \right)\left(
                     \begin{array}{cc}
                       -s_{\theta} & c_{\theta} \\
                       c_{\theta} & s_{\theta} \\
                     \end{array}
                   \right) = \left(
                             \begin{array}{c}
                               G'_{U} \\
                               h^0_4 \\
                             \end{array}
                           \right),\; \left(
                                        \begin{array}{c}
                                          S_{2} \\
                                          S_{1} \\
                                          S'_{3} \\
                                        \end{array}
                                      \right)= \left(
                                                 \begin{array}{ccc}
                                                   \frac{-c_{\alpha}}{\sqrt{2}} & \frac{s_{\alpha}}{\sqrt{2}} & -\frac{1}{\sqrt{2}} \\
                                                   \frac{-c_{\alpha}}{\sqrt{2}}& \frac{s_{\alpha}}{\sqrt{2}} & \frac{1}{\sqrt{2}} \\
                                                   s_{\alpha} & c_{\alpha} & 0 \\
                                                 \end{array}
                                               \right)
                                      \left(
                                                \begin{array}{c}
                                                  h^0_1 \\
                                                  h^0_2 \\
                                                  h^0_3 \\
                                                \end{array}
                                              \right),
        \label{ecpeh0}\eea
 where $s_{\alpha}=\sin\alpha$, $c_{\alpha}=\cos\alpha$ defining by
       \bea s_{\alpha}&=&\frac{4 \lambda_1 t^2-m^2_{h^0_1}/v_3^2}{\sqrt{2 \left( 2\lambda_1-f/v_3\right)^2 t^2+\left(4 \lambda_1 t^2-m^2_{h^0_1}/v_3^2\right)^2}},\crn
        c_{\alpha}&=&\frac{\sqrt{2}\left( 2\lambda_1-f/v_3\right) t}{\sqrt{2 \left( 2\lambda_1-f/v_3\right)^2 t^2+
        \left(4 \lambda_1 t^2-m^2_{h^0_1}/v_3^2\right)^2}}. \label{al1}\eea
\end{itemize}
In the limit   $t\ll1$ the expression of the lightest neutral even-CP Higgs is
$$ m^2_{h^0_1}\simeq v^2_1\left[ 4 \lambda_1-\frac{(\lambda_{12}-f/v_3)^2}{\lambda_2}\right],$$
where both $\lambda_1$ and $\lambda_2$ must be positive to guarantee the vacuum stability of the potential (\ref{331LHNpo}).
This Higgs is easily identified with the SM-like Higgs observed by LHC.
 \subsection{Couplings for LFV decay of the SM-like Higgs and the amplitude}
From the detailed discussions on the particle content of the 3-3-1LHN,  the couplings of SM-like Higgs needed for calculating LFVHD are collected   in the table \ref{smh0coupl}.

\begin{table}[h]
\scalebox{0.82}{
 \begin{tabular}{|c|c|c|c|}
\hline
 Vertex & Coupling & Vertex &Coupling \\
\hline
$\bar{N}_a e_bH_1^{+}$ & $-i\sqrt{2}V^{L*}_{ba}\left(\fr{m_{e_b}}{v_1}c_\theta P_R+\fr{m_{N_a}}{v_3}s_\theta P_L\right)$ & $\bar{e}_a N_bH_1^{-}$ &$-i\sqrt{2}V^{L}_{ba}\left(\fr{m_{e_b}}{v_1}c_\theta P_L+\fr{m_{N_a}}{v_3}s_\theta P_R\right)$ \\
\hline
$\bar{\nu}_ae_bH_2^{+}$&$-iU^{L*}_{ba}\left( \dfrac{m_{e_b}}{v_1}P_R+\dfrac{m_{\nu_a}}{v_2}P_L\right)$&$\bar{e}_b\nu_aH_2^{-}$&$-iU^{L}_{ab}\left( \dfrac{m_{e_b}}{v_1}P_L+\dfrac{m_{\nu_a}}{v_2}P_R\right)$\\
\hline
$\bar{N}_aN_ah_1^0$&$\fr{-i m_{N_a s_\al}}{v_3}$&$\bar{e}_ae_ah_1^0$&$\fr{im_{e_a}}{v_1}\fr{c_\al}{\sqrt{2}}$\\
\hline
$\bar{N}_ae_bV_\mu^{+}$&$\fr{ig}{\sqrt{2}}V^{L*}_{ba}\ga^\mu P_L$&$\bar{e}_bN_aV_\mu^{-}$&$\fr{ig}{\sqrt{2}}V^{L}_{ab}\ga^\mu P_L$\\
\hline
$\bar{\nu}_ae_bW_\mu^{+}$&$\fr{ig}{\sqrt{2}}U^{L*}_{ba}\ga^\mu P_L$&$\bar{e}_b\nu_aW_\mu^{-}$&$\fr{ig}{\sqrt{2}}U^{L}_{ab}\ga^\mu P_L$\\
\hline
$W^{\mu+}W_\mu^{-}h_1^0$&$-igm_W c_\al $&$V^{\mu+}V_\mu^{-}h_1^0$&$\dfrac{igm_V}{\sqrt{2}}(\sqrt{2}s_\al c_\theta-c_\al s_\theta)$\\
\hline
$h_1^0H_1^{+}V^{\mu-}$&$\dfrac{ig}{2\sqrt{2}}(c_\al c_\theta+\sqrt{2}s_\al s_\theta)(p_{h_1^0}-p_{H_1^{+}})_\mu$&$h_1^0H_1^{-}V^{\mu+}$&$\dfrac{ig}{2\sqrt{2}}(c_\al c_\theta+\sqrt{2}s_\al s_\theta)(p_{H_1^{-}}-p_{h_1^0})_\mu$ \\
\hline
$h_1^0 H_1^{+}H_1^{-}$&$-i v_3 \lambda_{h^0H_1H_1} $&$h_1^0H_2^{+}H_2^{-}$&$ -iv_1 \left[-2\sqrt{2}c_\alpha\lambda_1 +\frac{s_\al v_3 \la_{12} +s_\al f}{v_1}\right]$\\
\hline
$\bar{\nu}_a\nu_ah_1^0$&$\fr{im_{\nu_a}}{v_2}\fr{c_\al}{\sqrt{2}}$&$h_1^0H_2^{\pm}W^{\pm}_\mu$ &0\\
\hline
\end{tabular}}
\caption{Couplings relating with LFV of  SM-like Higgs decays in the 3-3-1LN model, where $\lambda_{h^0H_1H_1}=s_\al c^2_\theta\lambda_{12}+2s^2_\theta\lambda_2-\sqrt{2}(2c_\al c^2_\theta\lambda_1 +s^2_\theta \lambda_{12}) t_{\theta}-c_\theta s_\theta\frac{ f}{v_3}\sqrt{2}$. Here we only consider the couplings the unitary gauge.} \label{smh0coupl}
\end{table}
Matching the Feynman rules in the figure \ref{FeynRule}, we have the specific relations among the vertex parameters and  the couplings in the 3-3-1LHN, namely for the exotic leptons
\bea
a_1 &\rightarrow&  c_\theta,\;  a_2\rightarrow a_3 = s_\theta, \; v_1 = \fr{2m_V}{g}s_\theta,  \;  v_2\rightarrow v_3 = \fr{2m_V}{g}c_\theta, \crn
\fr{a_1}{v_1} &=& \fr{g}{2m_V}\fr{c_\theta}{s_\theta}, \; \fr{a_3}{v_3}=\fr{g}{2m_V}\fr{s_\theta}{c_\theta}, \; \fr{a_1a_3}{v_1v_3}=\fr{g^2}{4m^2_V},  \label{aijNlepton}
\eea
and the active neutrinos,
\be  a_1,a_2\rightarrow 1, \hs v_1,v_2\rightarrow v_1=v_2=\frac{v}{\sqrt{2}} =\frac{\sqrt{2}m_W}{g},\hs \frac{a_1}{v_1}=\frac{a_2}{v_2}=\frac{g}{\sqrt{2}m_W}. \label{aijactnu}\ee
 The expression  of $\Delta_L$ is separated into two parts, namely
\bea  \Delta^{N}_L &=& \sum_{a}V_{1a}^LV_{2a}^{L*}  \frac{1}{64\pi^2\sqrt{2}}
\left[\frac{}{}2 g^3\left(-c_{\alpha}s_{\theta}+ \sqrt{2}s_{\alpha}c_{\theta}\right)  \times E^{FVV}_{L}(m_{N_a},m_V) \right.\crn
&&+(- 2 g^2) \left( c_{\alpha}c_{\theta}+ \sqrt{2}s_{\alpha}s_{\theta}\right)\times E^{FVH}_{L} (a_1,a_3,v_1,v_3,m_{N_a},m_V,m_{H^{\pm}_1}) \crn
&&+ (-2g^2 )\left( c_{\alpha}c_{\theta}+ \sqrt{2}s_{\alpha}s_{\theta}\right)\times E^{FHV}_{L} (a_1,a_3,v_1,v_3,m_{N_a},m_V,m_{H^{\pm}_1}) \crn
&&+\left(- 4\sqrt{2}\lambda_{h^0H_1H_1}\right) \times E^{FHH}_{L} (a_1,a_2,v_1,v_2,m_{\nu_a},m_{H^{\pm}_2})\crn
&&+\frac{g^3s_{\alpha}\sqrt{2}}{c_{\theta}}\times E^{VFF}_L(m_V,m_{\nu_a})\crn
&&+\left(- 8\sqrt{2}s_{\alpha}\right)   E^{HFF}_{L} (a_1,a_3,v_1,v_3,m_{\nu_a},m_{H^{\pm}_1}) \crn
&&\left. +\frac{-g^3c_{\alpha}}{s_{\theta}}\times E^{FV}_L(m_V,m_{N_a})\right.\crn
&&+\left. 8c_{\alpha}\times E^{FH}_{L} (a_1,a_3,v_1,v_3,m_{N_a},m_{H^{\pm}_1})\right]  \label{NdeltaL1}\eea
from neutral exotic leptons and
\bea  \Delta^{\nu}_L &=&  \sum_{a}U_{1a} U_{2a}^{*} \frac{1}{64\pi^2}\left[  \frac{}{} \left(-2g^3c_{\alpha}\right)E^{FVV}_{L}(m_{\nu_a},m_W)\right. \crn
&&+(-4  \lambda_{h^0H_2H_2})\times E^{FHH}_{L} (a_1,a_2,v_1,v_2,m_{\nu_a},m_{H^{\pm}_2})\crn
&&+\left(-g^3c_{\alpha}\right) E^{VFF}_L(m_W,m_{\nu_a})\crn
&& +  (2\sqrt{2}c_{\alpha}) E^{HFF}_{L} (a_1,a_2,v_1,v_2,m_{\nu_a},m_{H^{\pm}_2}) \crn
&&\left. +\left(-g^3c_{\alpha}\right) E^{FV}_L(m_V,m_{\nu_a}) \right.\crn
&&+\left.  (2\sqrt{2}c_{\alpha}) E^{FH}_{L} (a_1,a_2,v_1,v_2,m_{\nu_a},m_{H^{\pm}_2})\right].  \label{nudeltaL1}\eea
Similarly for the $\Delta_R$ we have
 \be \Delta^N_R =  \Delta^N_L (E_L\rightarrow E_R), \hs \Delta^{\nu}_R =  \Delta^{\nu}_L (E_L\rightarrow E_R). \label{deltaR1}\ee
Before going to the numerical calculation we remind that the divergent cancellations  in two separate sectors of neutrinos and exotic leptons are presented precisely in the second subsection of the appendix \ref{Camlitude}.

\section{\label{num} Numerical investigation}
\subsection{Setup parameters}
In the  model under consideration, the new parameters we pay attention to are the  $SU(3)_L$ scale $v_3$, the mass of the lightest active neutrino, masses of the three neutral heavy leptons, Higgs masses and  mixing parameters of leptons and Higgses. The Higgs  part relates with the Higgs self-couplings  in the scalar potential: $ \lambda_1,\; \lambda_2,\;\lambda_{12}$ and $f$. The  first two free parameters we choose  are the $v_3$ and mass of the $H_2$ given in (\ref{schiggs}). Then the $f$ parameter can be determined by
\be  f= \frac{m^2_{H_2}}{2v_3}. \label{fre}\ee
 Another parameter that can be fixed is the mass of the neutral SM-like Higgs \cite{comHiggs} with the value of about $m_{h_1^0}=125.1$ GeV. Note that two Higgs masses $ m^2_{h^0_1}$ and $m^2_{h^0_2}$ shown in  (\ref{h0mass}) are roots of the equation $x^2+ax+b=0$, where $-a=m^2_{h^0_1}+m^2_{h^0_2}=v_3^2\left( 4\lambda_1 t^2+2\lambda_2+t^2f/v_3\right)$ and $b=m^2_{h^0_1}m^2_{h^0_2}=2v_1^2v_3^2\left[ 2\lambda_1\times\left(2\lambda_2+t^2f/v_3\right)-\left( \lambda_{12}-f/v_3\right)^2\right]$.  This means that $m^4_{h^0_1}+a\times m^2_{h^0_1}+b=0$, giving a  relation among $\lambda_{2}$,  $\lambda_1$ and $\lambda_{12}$:
\be \lambda_2=\frac{t^2_{\theta}}{2}\left( \frac{m^2_{h_1^0}}{v_1^2}-\frac{m^2_{H_2}}{2 v_3^2}\right)-  \frac{\left(\lambda_{12}-m^2_{H_2}/2v_3^2\right)^2}{-4\lambda_1+m^2_{h_1^0}/v_1^2}.\nn \label{lambda2}\ee
Because the $\lambda_1$, $\lambda_2$ and $\lambda_{12}$ are factors of quartic terms in the Higgs potential (\ref{331LHNpo}), they must satisfy  the unbounded from below (UFB) conditions that guarantee the stability of the vacuums of the considering model.  According to the ref. \cite{UFBc}, these conditions are easily found as follows. Defining $\rho^{\dagger}\rho+\eta^{\dagger}\eta=h_1^2$ and $\chi^{\dagger}\chi=h_2^2$, the quartic part of the Higgs potential (\ref{331LHNpo}) has form of $V_4= \lambda_1 (h_1^2)^2 +\lambda_{12} h_1^2h_2^2 +\lambda_2 (h_2^2)^2 $. In the basis $(h_1^2,h^2_2)$ the $V_4$ corresponds to the  $2\times2$ matrix that must satisfy the conditionally positive conditions as follows:
\be  \lambda_1 >0,\hs \lambda_2>0, \hs  \mathrm{and}\; \frac{\lambda_{12}}{2}+\sqrt{\lambda_1\lambda_2}\geq0. \label{UFBc1}\ee
In our calculation, apart from positive $\lambda_1$ and $\lambda_2$ we will choose $\lambda_{12}>0$ so that all conditions given in (\ref{UFBc1}) are always satisfied.

To identify  $h^0_1$ with the SM Higgs, the $h^0_1$  must satisfy  new constrains from LHC, as  discussed in \cite{LHCSMh0}. Namely, the mixing angle $\alpha$ of neutral Higgses, defined in (\ref{al1}), should be constrained from the $h^0_1W^{+}W^{-}$ coupling.  Following  \cite{LHCSMh0}  the  we can identify that  $-c_{\alpha}\equiv 1+\epsilon_{W}$ where $ \epsilon_{W}=-0.15\pm 0.14$ is the universal fit for the SM Higgs. This results the constraint of $c_{\alpha}$ as
 \be -0.99\leq c_{\alpha}\leq -0.71.\label{Ufit} \ee
  By canceling a factor of $t$ in (\ref{al1}),  we have a simpler expression
$$c_{\alpha}=\frac{ \sqrt{2}\left( 2\lambda_1-\frac{m_{H_2}^2}{2v_3^2}\right)}{\sqrt{2\left( 2\lambda_1-\frac{m_{H_2}^2}{2v_3^2}\right)^2+ t^2\left(4\lambda_1-m^2_{h^0_1}/v_1^2 \right)^2}},$$
which shows that $ c_{\alpha}<0$ when $ m_{H_2}> 2 v_3\sqrt{\lambda_1}$ and  $ c_{\alpha}\rightarrow -1$ when $t \ll 1$. The lower constraint of $c_{\alpha}$ in (\ref{Ufit}) gives a very interesting relation among $\lambda_1,~v_3$ and $m_{H_2}$, namely $m^2_{H_2}$ can be written as
\be  m^2_{H_2}=v_3^2\left[ 4\lambda_1 + \left| 4\lambda_1-\frac{m^2_{h^0_1}}{v_1^2}\right|\times \frac{\sqrt{2}|c_{\alpha}|}{\sqrt{1-c^2_{\alpha}}}\times \frac{v_1}{v_3}\right]. \label{mh2}\ee
If  the lower constraint in (\ref{Ufit}) is not considered, $m^2_{H_2}$ can be arbitrary large when $|c_{\alpha}|\rightarrow 1$. In contrast, the constraint   (\ref{Ufit}) gives a consequence  $\frac{\sqrt{2}|c_{\alpha}|}{\sqrt{1-c^2_{\alpha}}} \sim \mathcal{O}(1)$. Combining with  $m^2_{h^0_1}/v_1^2\simeq 0.52$, we obtain a rather strict relation $m_{H_2}\simeq 2v_3\sqrt{\lambda_1}$ if $v_3 \gg v_1 \simeq246/\sqrt{2}$ GeV and $\lambda_1$ is large enough. On the other hand, this relation will not hold if the custodial symmetry assumed in the Higgs potential (\ref{331LHNpo}) is only an approximation. Hence in the numerical calculation, for the general case we will first investigate the LFVHD without the constraint (\ref{Ufit}). This constraint will be discussed in the final.

Regarding to the parameters of active neutrinos we use the recent results of experiment. In particularly, if  the mixing parameters in the active neutrino sector are parameterized by
\bea U (\theta_{12},\theta_{13},\theta_{23})&=&\left(
           \begin{array}{ccc}
             1 & 0 & 0 \\
             0 & \cos\theta_{23} & \sin\theta_{23} \\
             0 & -\sin\theta_{23} & \cos\theta_{23} \\
           \end{array}
         \right)
\left(
           \begin{array}{ccc}
             \cos\theta_{13} & 0 & \sin\theta_{13} \\
             0 & 1 & 0 \\
             -\sin\theta_{13} & 0 & \cos\theta_{13} \\
           \end{array}
         \right)\crn
         &\times&
\left(
           \begin{array}{ccc}
             \cos\theta_{12} & \sin\theta_{12} & 0 \\
             -\sin\theta_{12} & \cos\theta_{12} & 0 \\
             0 & 0 & 1 \\
           \end{array}
         \right). \label{mixingpar}\eea
 Because $U^{L}$  has a small deviation from the  well-known neutrino mixing  matrix $U^{MNPS}$ so we ignore this deviation \cite{numixing}. We will use the best-fit  values of neutrinos oscillation parameters given in \cite{nuosc},
\bea \Delta m^2_{21}&=& 7.60\times 10^{-5}\;\mathrm{ eV^2},\hs  \Delta m^2_{31}= 2.48\times 10^{-3}\; \mathrm{eV^2},\crn
\sin^2\theta_{12}&=&0.323,\; \sin^2\theta_{23}=0.467,\; \sin^2\theta_{13}=0.0234,\label{nuosc}\eea
and mass of the lightest neutrino will be chosen in range $10^{-6}\leq  m_{\nu_1}\leq 10^{-1}$ eV, or  $10^{-15}\leq m_{\nu_1}\leq 10^{-10}$ GeV. This range  satisfies the condition $\sum_{b}m_{\nu_b}\leq 0.5$ eV obtained from the cosmological observable. The remain two neutrino masses are $m^2_{\nu_b}=m^2_{\nu_1}+\Delta m^2_{\nu_{b1}}$. We note that the above case corresponds to the normal hierarchy of active neutrino masses. In the 3-3-1LHN, the inverted case gives the same result so we do not consider here.

The mixing matrix of the exotic leptons is also parameterized according to (\ref{mixingpar}). In particularly it  is unknown and defined as $V^L \equiv U^L(\theta^N_{12}, \theta^N_{13},\theta^N_{23})$. If all $\theta^N_{ij}=0$, all contributions from exotic leptons to $\Delta_{L,R}$ will be exactly zero. In the numerical computation, we consider only the cases of maximal mixing in the exotic lepton sector, i.e.  each $\theta^N_{ij}$  gets  only the value of $\pi/4$ or zero.  There are  three interesting  cases: i) $\theta^N_{12}=\pi/4$ and $\theta^N_{13}=\theta^N_{23}=0$; ii) $\theta^N_{12}=\theta^N_{13}=\theta^N_{23}=\pi/4$; and iii) $\theta^N_{12}=\theta^N_{13}=\pi/4$ and $\theta^N_{23}=-\pi/4$.  The other cases just change  minus signs in the total amplitudes, and do not change the final results of LFVHD branching ratios. For example the mixing matrix of first case is
 \bea V^L=U(\pi/4,0,0)= \left(
           \begin{array}{ccc}
             \frac{1}{\sqrt{2}} &  \frac{1}{\sqrt{2}} & 0 \\
             - \frac{1}{\sqrt{2}} &  \frac{1}{\sqrt{2}} & 0 \\
             0 & 0 & 1 \\
           \end{array}
         \right).\label{maxE1}\eea
Our numerical investigation will pay attention to the first case, where the third exotic lepton does not contribute to the LFVHD decays. The two other cases are easily deduced from this investigation.

From the above discussion, we chose the following unknown parameters as free parameters: $v_3$, $m_{H_2}$, $ \lambda_1,\;\lambda_{12}$, $m_{\nu_1}$ and $m_{N_a}$ ($a=1,2,3$).  The vacuum stability of the potential (\ref{331LHNpo}) results  the consequence $\lambda_{1,2}>0$. In order to be consistent with the perturbativity property of the theory,  we will choose $ \lambda_{1}, |\lambda_{12}| < \mathcal{O}(1)$. The numerical check shows that the LFVHD branching ratio depends weakly on the changes of  these Higgs self-couplings   in this range. Therefore  we will fix $\lambda_1=\lambda_{12}=1$ without  loss of generality. These values of $\lambda_1$ and $\lambda_{12}$  also satisfy all UFB conditions (\ref{UFBc1}). In addition, the Yukawa couplings in the Yukawa term (\ref{Ylepton1}) should have a certain upper bound, for example  in order to be consistent with the perturbative unitarity limit \cite{Ubound1}. Because the vev $v_3$ generates masses for exotic leptons from the Yukawa interactions (\ref{mterm}), following  \cite{MainHll} we assume  the  upper bound of the lepton masses as follows
\be\left| \frac{m_{N_a}}{v_3}\right|^2 \leq\left|\frac{y^N_{ij}}{\sqrt{2}}\right|^2<3\pi. \label{Yupper}\ee
  After investigating the dependence of the LFVHD on the Yukawa couplings  through the ratio $\frac{m_{N_a}}{v_3}$ we will fixed $m_{N_2}/v_3=0.7 $ and 2 corresponding to the two  cases of lower and larger than 1 of the Yukawa couplings.

  Unlike the assumption in \cite{snu331} where $f=v_3/2$, we treat $f$ as a free parameter relating with $m_{H_2}$ by the equation (\ref{fre}),  so the condition  of candidates of DM may be changed. We stress that the correlation between  $m_{H_2}$ and $v_3$ is very important to get maximal values of LFVHD branching ratio.  The singly charged Higgs bosons have been being searched at LHC, namely  the decays $H^+\rightarrow c\bar{s}$ or $ H^{\pm}\rightarrow W Z$  with  ATLAS  \cite{lowerboundHpm}, and decays to fermions with CMS \cite{singlyHCMS}. The ATLAS  gives a lower bound of 1 TeV while that from CMS is about 600 GeV. But in the 3-3-1LHN model, there is no coupling $H_1^{\pm}W^{\mp}Z$, while the coupling $H_2^{\pm}W^{\mp}Z$ is extremely small when $v_1=v_2$.  In addition,  only the $H_2$ decay has been searched by CMS so the lower bound of $m_{H_2}\geq 600$ GeV should be applied.   The value of $m_{H_2}$ should also satisfy $\frac{m_{H_2}}{v_3} \leq\mathcal{O}(1)$, resulting an upper bound depending on the $SU(3)_L$ scale.

   The value of $v_3$  should be consistent with  the lower bound of  $Z'$ mass from experimental searches \cite{Zpex}, addressing directly for 3-3-1 models  \cite{prtests331,zpbound}, where $m_{Z'}$ must be above 2.5 TeV. It is enough using an approximate relation of $m_{Z'}$ and $v_3$: $m_{Z'}^2\simeq g^2v_3^2 c^2_W/(3-4s^2_W)$ where $s_W=\sin\theta_W$ and  $c_W=\cos\theta_W$ with $\theta_W$ being the Weinberg angle. Then $v_3$ should be above 6 TeV.  For understanding the qualitative properties  of the LFVHD, our investigation will pay attention on the  range of  $4\mathrm{TeV}<v_3<10$ TeV.

To see the correlation between singly charged Higgses, the neutral leptons and the $v_3$,  the range of $m_{H_2}$ will be chosen as   $0.5\mathrm{TeV}< m_{H_2}<20$ TeV.  The default value of $m_{N_1}=2$ TeV is used. The value of $m_{N_2}$ is chosen later.

 The other well-known parameters are fixed \cite{PDG2014}: $W$ boson mass $m_{W}=80.385$  GeV, the weak-mixing angle value $s^2_{W}=0.231$, the fine-structure constant at the electroweak scale $ \alpha=e^2/4\pi=1/128$, the total decay width of the SM Higgs $\Gamma_H\simeq 4.07$ GeV.  The mass of this Higgs is fixed as  $m_H=125.09$ GeV. These two values are assumed to be the total decay width and mass of the SM-like Higgs considered in this work.

A main point that can distinguish the  LFVHD characteristics  in the 3-3-1 models with the other well-known models beyond SM, including the seesaw and SUSY models,  is the relation of new neutral lepton masses and the Yukawa couplings which directly relate to the LFVHD. In particular, because all neutral leptons  in 3-3-1LHN  receive masses from the Yukawa terms, so their masses must  be bounded from above because of the inequality (\ref{Yupper}) and a similar one for active neutrinos. This also implies that maximal values of exotic lepton masses depend on the $SU(3)_L$ scale $v_3$. While in the  seesaw models with new singlets right-handed neutrinos,  the mass terms of sterile neutrinos are mainly come from the private Majorana mass terms and no new Yukawa couplings appear. So the mass ranges of new sterile neutrinos may be very wide, even if their effects to the Yukawa couplings of the active neutrinos are included \cite{MainHll}. Similar, in the SUSY models, the appearance of the soft terms leads to the consequence that masses of new superpartners affecting to LFVHD are mainly come from these soft terms.  In conclusion, the study of LFVHD  in 3-3-1LHN can give some interesting information on Yukawa couplings of exotic leptons and the  $SU(3)_L$ scale $v_3$.
\subsection{Numerical result}
\begin{figure}[h]
 \centering
\begin{tabular}{cc}
\includegraphics[width=6.8cm]{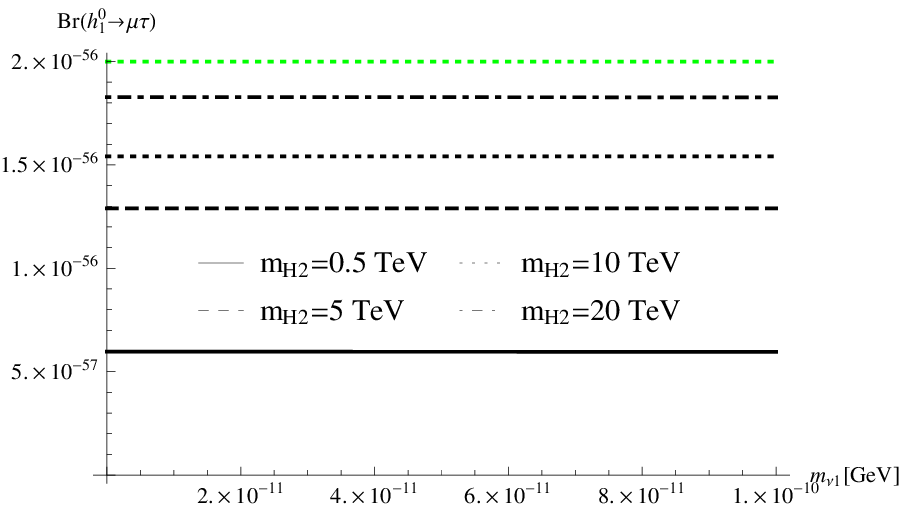}& \includegraphics[width=6.8cm]{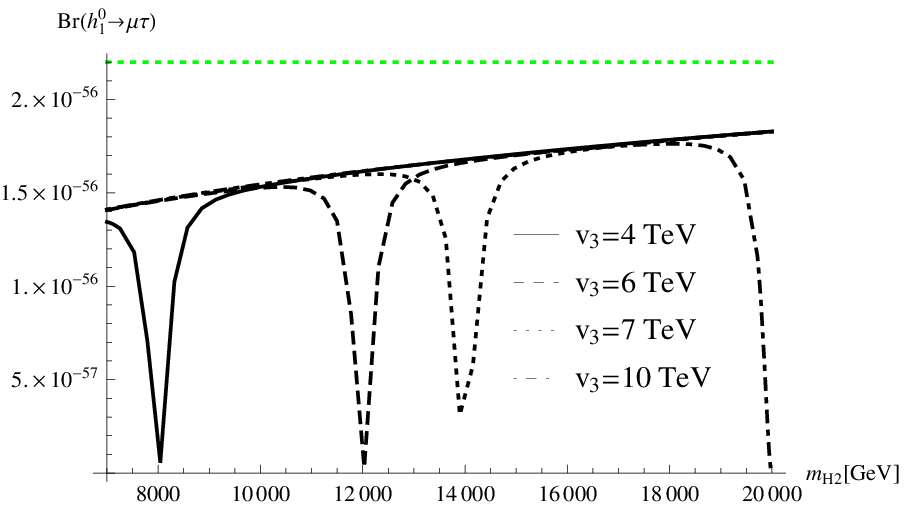}\\
 \end{tabular}
 \caption{ Branching ratio of LFVHD as function of $ m_{\nu_1}$ (left panel) or $m_{H_2}$ (right panel) where contributions are come from only active neutrinos in the loops.}\label{nubr}
\end{figure}

If the mixing parameters among all exotic leptons are zero or all of their masses are degenerate, then the contributions to the LFVHD  of these exotic leptons are zero, too. Then branching ratio of the LFVHD $h^0_1\rightarrow \mu\tau$ depends on only active neutrino sector, in which the mixing parameters as well as masses are almost known. The numerical results in this case are shown in the figure \ref{nubr}. The LFVHD does not depend on the value of the lightest active neutrino, but increases very slightly with the increasing  of $v_3$ and $m_{H_2}$. Because both values of $v_3$ and $m_{H_2}$  are  in the TeV scale, the contribution of the active neutrinos is extremely small compared  with the recent experimental sensitivity, so we can neglect it in the next calculation.

\begin{figure}[h]
  \centering
 \begin{tabular}{cc}
  \includegraphics[width=6.8cm]{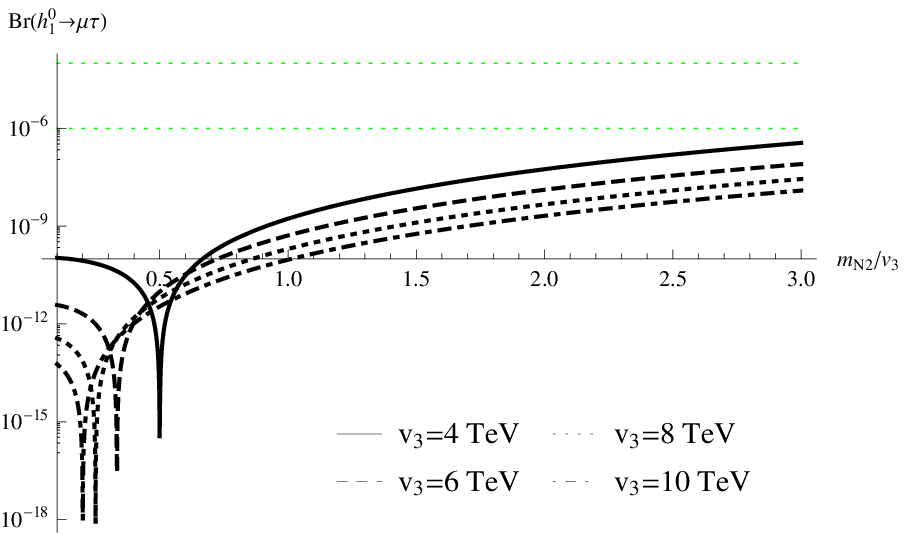}& \includegraphics[width=6.8cm]{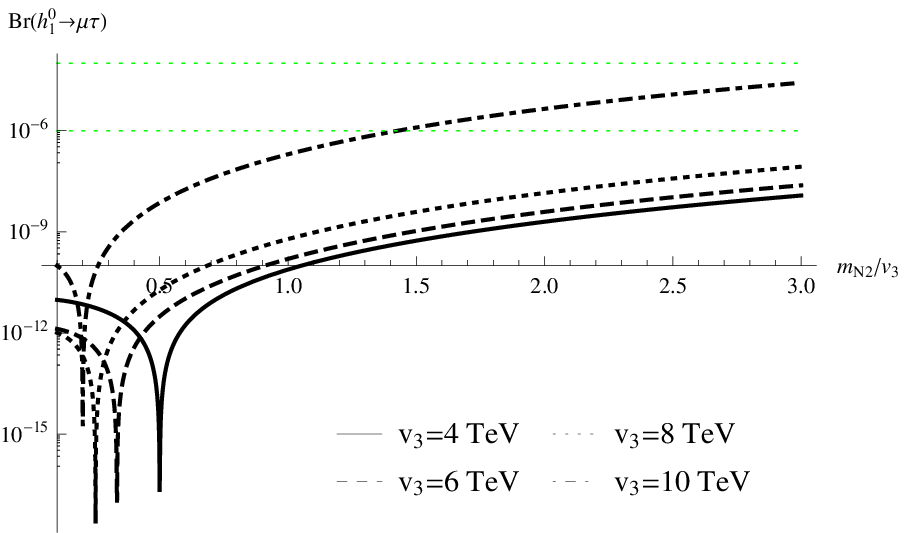}\\
  \end{tabular}
  \caption{Branching ratio of LFVHD as function of $m_{N_2}/v_3$, which is proportional to Yukawa couplings of exotic leptons,  $m_{H_2}=2$ (20) TeV in the left (right) panel. The upper green lines correspond to the value of $10^{-4}$.}\label{fYukawa}
\end{figure}

Now we begin considering the contribution of exotic leptons. Firstly  the dependence of the branching ratio of LFVHD on the Yukawa couplings, or the ratio of $m_{N_2}/v_3$, is  shown in the figure \ref{fYukawa}.  The branching ratio enhances rapidly with the increasing of the Yukawa couplings. In addition, the branching ratio is small, below $10^{-6}$,  with small $m_{H_2}=2$ TeV,  and rather large with larger $m_{H_2}$.  In particular for $m_{H_2}=20$ TeV, the branching ratio can reach $10^{-5}$.
 Both of the largest values in the two panels correspond to the largest values of the Yukawa couplings. The deep wells show the zero values of the LFVHD branching ratio when the two exotic lepton masses are exactly degenerate at the default value of $m_{N_1}=2$ TeV. For the small value of $m_{H_2}$, the small $v_3$ (the black line in the left panel) gives larger BR$(h^0_1\rightarrow\mu\tau)$.  In contrast, the larger values of $m_{H_2}$ and  $v_3$   (the dot-dash line in the right panel) give large  BR$(h^0_1\rightarrow\mu\tau)$. The one more interesting property is that the branching ratio seems to be unchanged with very small values of $m_{N_2}$, implies that the small  exotic lepton masses give small contribution the to LFVHD.  The constant values of LFVHD in the right-hands sides of the wells are from the contributions of $m_{N_1}=2$ TeV when $m_{N_2}$ is much smaller than $m_{N_1}$.

For qualitative   estimation, we have checked  $\Delta_{L,R}$ as functions of mass parameters as follows. We divide them into two parts: $\Delta_{L,R}=f(m_H,v_3,m_{N_a})+g(m_H,v_3,m_{N_a}) \times\ln(m^2_{N_a})$ and consider their behavior  when one of the parameters approaches zero or infinity. Note that the logarithm factors are very important because they can give very large contributions even with the very small values of $m_{N_a}$. For the exotic lepton masses, there are two interesting properties:
\be   \lim_{m_{N_a}\rightarrow0} g(m_H,v_3,m_{N_a})\ln(m^2_{N_a})=0\;  \mathrm{and}\;  \lim_{m_{N_a}\rightarrow\infty} g(m_H,v_3,m_{N_a}) \ln(m^2_{N_a})=\pm\infty,  \label{rlnma}\ee
with the assumption that all other parameters are fixed and the exotic lepton masses  do not have any upper bounds.  The first limitation explains why small exotic leptons give suppressed contributions to LFVHD. If  the upper bound of the Yukawa couplings, namely (\ref{Yupper}),  is  applied,  the value of the second limitation in (\ref{rlnma}) becomes zero. In the well-known classes of models such as the models with singlet right-handed neutrinos or the SUSY models, the upper bounds of  new lepton masses  or superpartner masses do not relate with the vevs of Higgses, because these masses are also come from other sources as the singlet mass terms or the soft terms.
So the  Br$(h^0_1\rightarrow\mu\tau)$  increases with increasing of the new mass scales \cite{MainHll}. Hence the upper bound of the LFVHD will result to the upper bound of these new mass scales. In contrast, in the frame work of the 3-3-1 models, the LFVHD  will give much of important information of  the Yukawa couplings of the exotic leptons.

As showed in the figure \ref{fYukawa}, the Br$(h^0_1\rightarrow\mu\tau)$ depends clearly on  $m_{N_2}/v_3$ whether this ratio is larger or smaller than 1.  From now we will consider two fixed values of $m_{N_2}/v_3=0.7$ and 2, without  any inconsistence in the final results.
\begin{figure}[h]
  \centering
 \begin{tabular}{cc}
  \includegraphics[width=6.8cm]{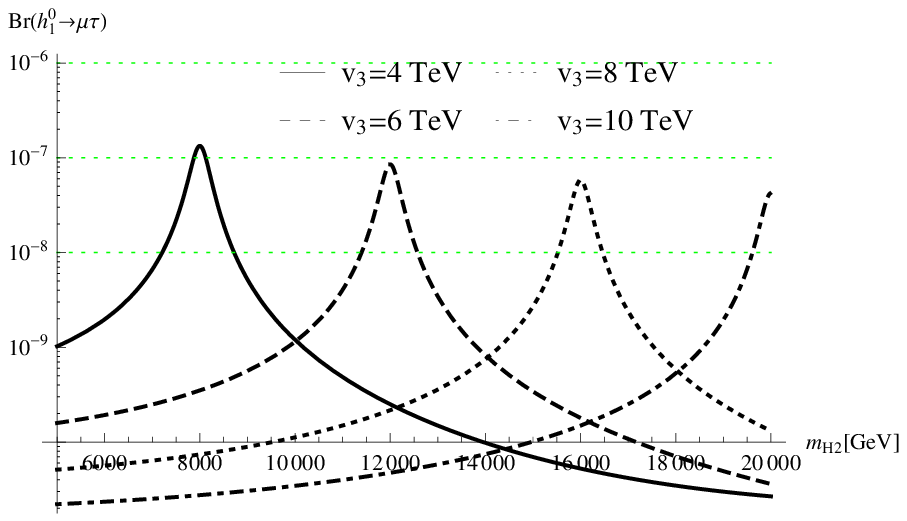}& \includegraphics[width=6.8cm]{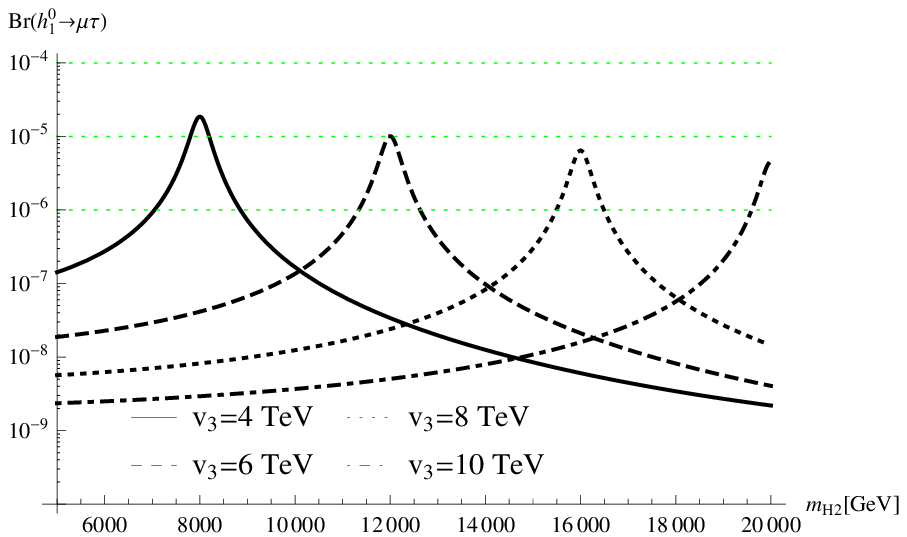}\\
  \end{tabular}
  \caption{Branching ratio of LFVHD as function of $m_{H_2}$,  $m_{N_2}/v_3=0.7$ (2) in the left (right) panel. }\label{fmH2}
\end{figure}

The figure \ref{fmH2} shows the dependence of  LFVHD on the mass of $m_{H_2}$. The first property we can see is that the LFVHD branching ratio always has an upper bound that decreases with  increasing $v_3$. In other word, it has an maximal value depending strictly on the constructive correlation of $v_3$ and $m_{H_2}$. But if  the Yukawa couplings are small, this maximum  seems never reach the value of $10^{-6}$. The case of the large Yukawa couplings is more interesting because maximal LFVHD can be asymptotic $10^{-5}$, provided that  $v_3$ is small enough, see the right panel.
\begin{figure}[h]
  \centering
 \begin{tabular}{cc}
  \includegraphics[width=6.8cm]{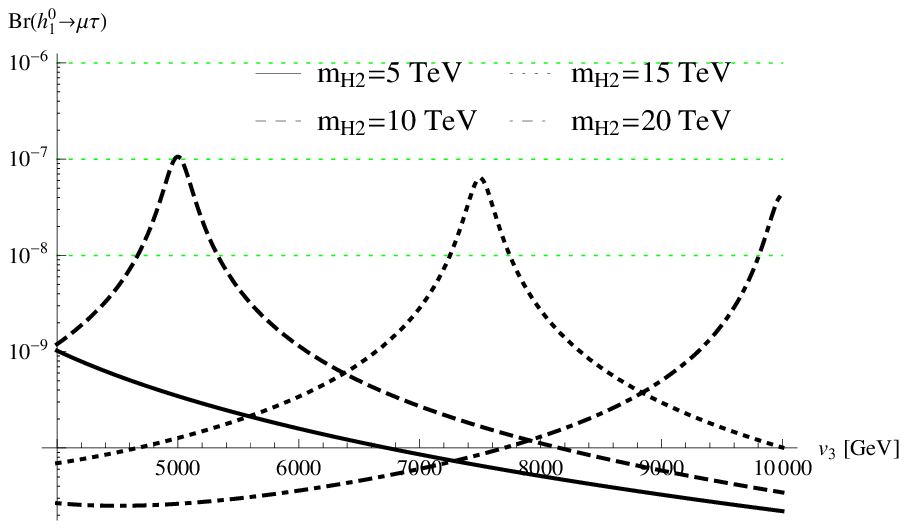}& \includegraphics[width=6.8cm]{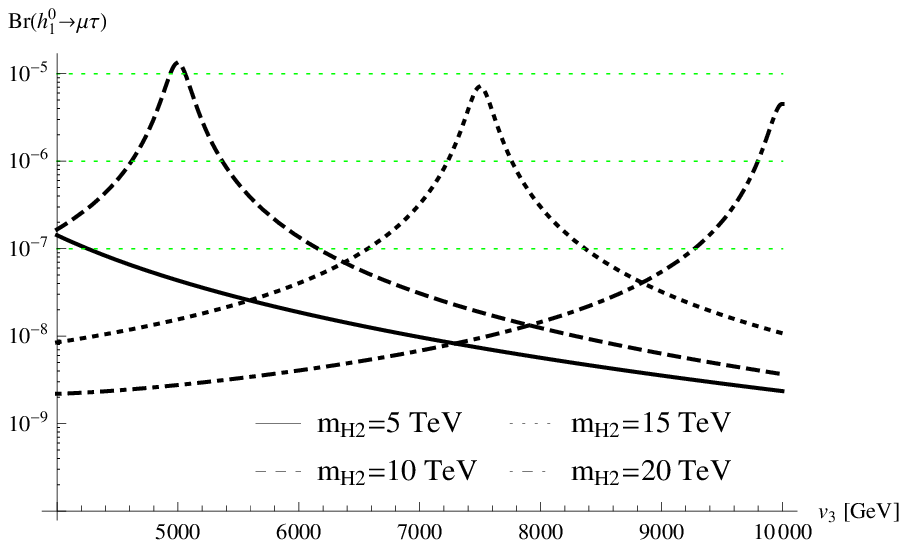}\\
  \end{tabular}
  \caption{Branching ratio of LFVHD as function of $v_3$,  $m_{N_2}/v_3=0.7$ (2) in the left (right) panel. }\label{fv3}
\end{figure}

The effects of $v_3$ on LFVHD are shown in the figure \ref{fv3}. Again we can see that the maximal values can reach $10^{-7}$ and  $10^{-5}$ for respective small and large Yukawa couplings.

Combining  both figures \ref{fmH2} and \ref{fv3}, we conclude that  the construction correlation of $m_{H_2}$ and $v_3$ is the necessary condition for maximal peaks and the appearance of vertices are independent with Yukawa couplings.  But the  maximal values of LFVHD branching ratio depend directly on the amplitudes of the Yukawa couplings and can reach $10^{-5}$.
\begin{figure}[h]
  \centering
 \begin{tabular}{cc}
  \includegraphics[width=6.6cm]{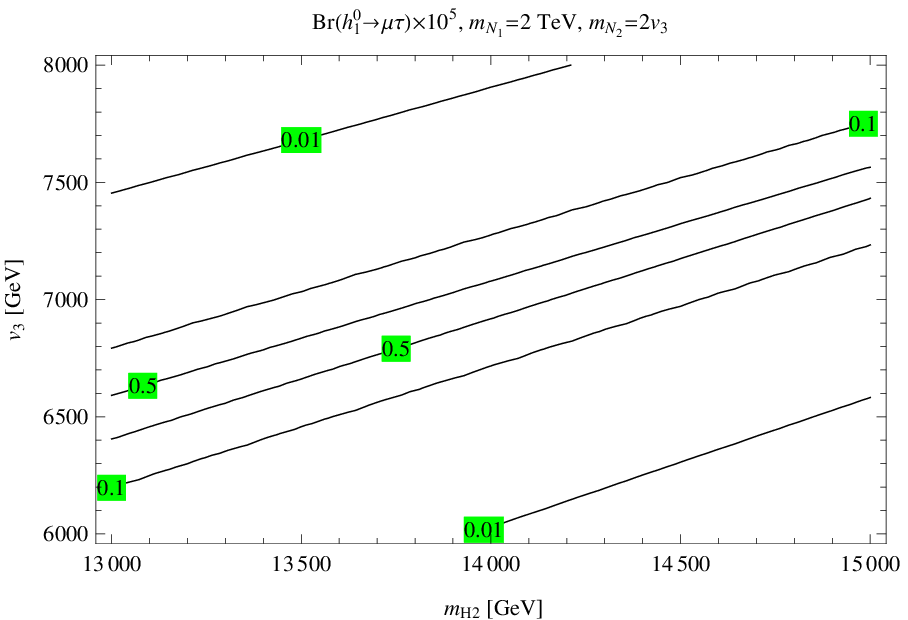}& \includegraphics[width=6.6cm]{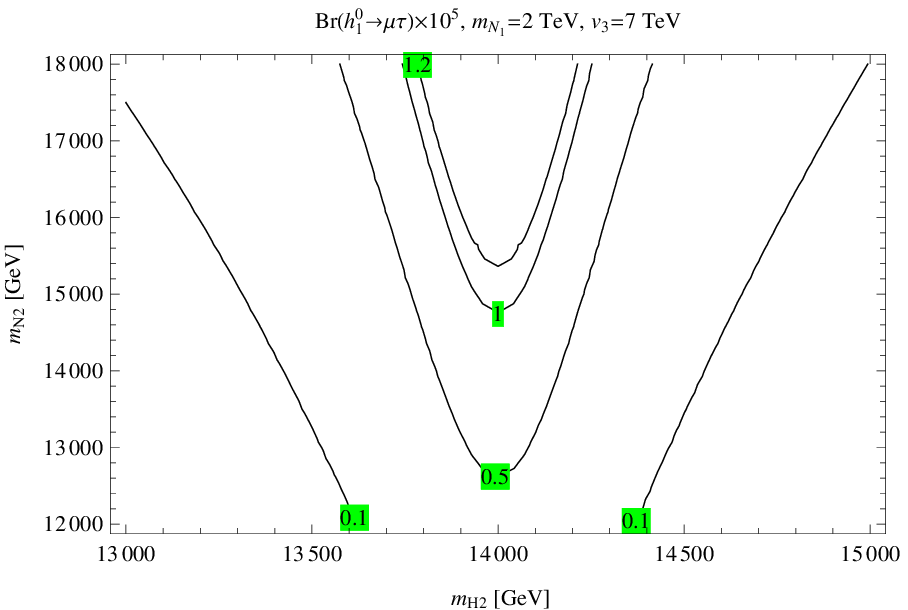}\\
  \end{tabular}
  \caption{Contour plots of LFVHD as function of $v_3$ and $m_{H_2}$  in the left (right) panel. }\label{contourv3H}
\end{figure}

The figure \ref{contourv3H} represents some particular regions of the  parameter space to get the large values of LFVHD Br$(h^0_1\rightarrow \mu\tau) $. Especially the values larger than $10^{-5}$ are  the maximal values of LFVHD that the  3-3-1LHN can predict when the lower bound of $v_3$ is 6 TeV. In addition,  the left panel shows the case of $m_{N_2}/v_3=2$, the parameters satisfying Br$(h^0_1\rightarrow\mu\tau)\geq 0.5\times10^{-6}$ is very narrow, implies a very strict relation of $v_3$ and $m_{H_2}$ if this large amount  of the branching ratio is observed. The right panel shows the dependence of Br$(h^0_1\rightarrow\mu\tau)$ on the Yukawa couplings and $m_{H_2}$ with $v_3=7$ TeV. Clearly, the maximal peak of LFVHD corresponds to $m_{H_2}\simeq 14$ TeV and does not depend on the Yukawa couplings. But the maximal values do, in this case Br$(h^0_1\rightarrow\mu\tau)\geq 0.5\times10^{-5}$ if only $m_{N_2}\geq 14.5$ TeV. Furthermore, the region having  Br$(h^0_1\rightarrow\mu\tau)\geq 0.5\times10^{-5}$ opens wider with larger Yukawa couplings.

Finally, we should pay attention to the case satisfying the constraint of universal Higgs fit (\ref{Ufit}). In the above numerical investigation, we have fixed $\lambda_1=1$, which corresponds to  $m_{H_2}\simeq 2v_3 \sqrt{\lambda_1}=2 v_3$ satisfying  the constraint. It is very interesting that  all maximal peaks of LFVHD appearing in the numerical calculations  correspond to this relation among $m_{H_2},v_3$ and $\lambda_1$. Therefore the universal Higgs fit confirms more strongly that  the 3-3-1LHN predicts the  large branching ratios of LFVHD.

\section{Conclusion}
For studying the LFVHD in the 3-3-1LHN model, we have introduced form factors expressing the one-loop contributions corresponding to relevant Feynman diagrams in the unitary gauge. We have checked that the total contribution is finite, all of the divergences appearing in particular diagrams cancel among one to another.   Although the above form factors  are calculated for the 3-3-1LHN, they can be applied for other 3-3-1 models and in general for many other models beyond the SM with the same class of particles.  In numerical investigation the LFVHD in the case of maximal mixing between the first two exotic neutral leptons, we  find that the branching ratio Br$(h^0_1\rightarrow\mu\tau)$ depends the mostly  on Yukawa couplings of neutral exotic leptons and  the $SU(3)_L$ scale $v_3$.  For small $y^{N}_{ij}\simeq 1$, equivalently $m_{N_2}/v_3\simeq0.7$, this  branching ratio is always lower than $10^{-6}$, and even that of about $10^{-7}$,  the parameter space is very narrow. In contrast, with large Yukwa couplings, for example  $y^{N}_{ij}\simeq 2\sqrt{2}$ or $m_{N_2}/v_3\simeq2$,the largest LFVHD branching ratio
can reach $10^{-5}$ and does not depend on the small values of $m_{N_1}$.  These largest values do also depend on the charged Higgs masses and the $v_3$, thought these seem not as strongly as the Yukawa couplings. The values above $10^{-5}$ can be found in large region of parameter space with small $v_3$.  With the large $v_3$, this region is very small, implying some strict relation between parameters of exotic lepton masses, charged Higgs masses and the $SU(3)_L$ scale $v_3$.  The  relation arises from  the present  of both the custodial symmetry in the Higgs potential and the constraint from the universal fit of the Higgs property observed by LHC.  This will give interesting information of the 3-3-1LHN model if the  LFVHD branching ratio is  discovered by experiments at the value of $10^{-5}$ or larger.  Our calculation also indicates that  only 3-3-1 models with new heavy leptons, such as \cite{newHeavyL},  can predict large LFVHD. So when calculating the LFVHD in  SUSY versions, the non-SUSY contributions must be included. In contrast, the 3-3-1 models with light leptons \cite{newLihgtL} give suppressed signals of LFVHD, and the SUSY-contributions in \cite{SUSY331} are dominant.

\section*{Acknowledgments}
L.T. Hue would like to thank Le Duc Ninh and Phan Hong Khiem for helpful discussion on divergent cancellation and  the  idea of Le Duc Ninh about using Form  to compare the one-loop formulas as functions of PV-functions.  This research is funded by Vietnam  National Foundation for Science and Technology Development (NAFOSTED) under grant number 103.01-2015.33.
\appendix

\section{\label{mainte}Master integrals for one-loop integral calculation}
\subsection{Master integrals}
The calculation in this section relates with one-loop diagrams
 in the figure \ref{nDiagram}.  We introduce the notations $D_0=k^2-M_0^2+i\delta$, $D_1=(k-p_1)^2-M_{1}^2+i\delta$ and $D_2=(k+p_2)^2-M_2^2+i\delta$, where $\delta$ is  infinitesimally a  positive real quantity. The scalar integrals are defined as
 \bea
A_{0}(M_i)
 &=&\frac{\left(2\pi\mu\right)^{4-D}}{i\pi^2}\int \frac{d^D k}{D_i}, \hs
 B^{(1)}_0 \equiv\frac{\left(2\pi\mu\right)^{4-D}}{i\pi^2}\int \frac{d^D k}{D_0D_1},\crn
 B^{(2)}_0 &\equiv &\frac{\left(2\pi\mu\right)^{4-D}}{i\pi^2}\int \frac{d^D k}{D_0D_2}, \hs
  B^{(12)}_0 \equiv \frac{\left(2\pi\mu\right)^{4-D}}{i\pi^2}\int \frac{d^D k}{D_1D_2},\crn
 C_0&\equiv&  C_{0}(M_0,M_1,M_2) =\frac{1}{i\pi^2}\int \frac{d^4 k}{D_0D_1D_2},
 \label{scalrInte}\eea
 where $i=1,2$.
   In addition, $D=4-2\epsilon \leq 4$ is the dimension of the integral.  The notations $~M_0,~M_1,~M_2$ are masses  of virtual particles in the loops. The momenta satisfy conditions: $p^2_1=m^2_{1},~p^2_2=m^2_{2}$, and $(p_1+p_2)^2=m^2_{h^0}$. The tensor integrals are
 \bea
A^{\mu}(p_i;M_i)
 &=&\frac{\left(2\pi\mu\right)^{4-D}}{i\pi^2}\int \frac{d^D k\times k^{\mu}}{D_i}=A_0(M_i)p_i^{\mu},\crn
 B^{\mu}(p_i;M_0,M_i)&=& \frac{\left(2\pi\mu\right)^{4-D}}{i\pi^2}\int \frac{d^D k\times
k^{\mu}}{D_0D_i}\equiv B^{(i)}_1p^{\mu}_i,\crn
 B^{\mu}(p_1,p_2;M_1,M_i)&=& \frac{\left(2\pi\mu\right)^{4-D}}{i\pi^2}\int \frac{d^D k\times
k^{\mu}}{D_1D_2}\equiv B^{(12)}_1p^{\mu}_1+B^{(12)}_2p^{\mu}_2,\crn
C^{\mu} &=&C^{\mu}(M_0,M_1,M_2)=\frac{1}{i\pi^2}\int \frac{d^4 k\times k^{\mu}}{D_0D_1D_2}\equiv  C_1 p_1^{\mu}+C_2 p_2^{\mu},\crn
 \label{oneloopin1}\eea
where $A_0$, $B^{(i)}_{0,1}$, $B^{(12)}_{i}$ and $C_{0,1,2}$   are PV- functions.  It is well-known that $C_{i}$ is finite while the remains are divergent. We define
\be \Delta_{\epsilon}\equiv \frac{1}{\epsilon}+\ln4\pi-\gamma_E+\ln\frac{\mu^2}{m_h^2}, \label{divt}\ee where $\gamma_E$ is the  Euler constant and $m_h$ is the mass of the neutral Higgs.  The divergent parts of the above scalar factors can be determined as
\bea  \mathrm{Div}[A_0(M_i)]&=& M_i^2 \Delta_{\epsilon}, \hs  \mathrm{Div}[B^{(i)}_0]= \mathrm{Div}[B^{(12)}_0]= \Delta_{\epsilon}, \crn
\mathrm{Div}[B^{(1)}_1]&=&\mathrm{Div}[B^{(12)}_1] = \frac{1}{2}\Delta_{\epsilon},  \hs  \mathrm{Div}[B^{(2)}_1] = \mathrm{Div}[B^{(12)}_2]= -\frac{1}{2} \Delta_{\epsilon}.  \label{divs1}\eea
We remind that the  finite parts  of  the PV-functions such as B-functions  depend  on the scale of $\mu$ parameter with the same coefficient of the divergent parts.

 The analytic formulas of the above PV-functions are:
 \be  A_0(M)=M^2\left( \Delta_{\epsilon}+ \ln\frac{m_h^2-i\delta}{M^2-i\delta}+1\right)\equiv M^2\Delta_{\epsilon}+a_0(M), \label{afac}\ee
 \be B^{(i)}_{0,1}= \mathrm{Div}[B^{(i)}_{0,1}]+ b^{(i)}_{0,1}, \hs   B^{(12)}_{0,1,2}= \mathrm{Div}[B^{(12)}_{0,1,2}]+ b^{(12)}_{0,1,2}, \label{B01i}\ee
 where
\bea b^{(i)}_0 &=&\ln(m_h^2-i\delta)-\int^1_0 dx \ln\left[x^2p_i^2-x(p_i^2+M^2_0-M^2_i)+M_0^2-i\delta\right],\crn
 b^{(12)}_0 &=& \ln(m_h^2-i\delta)-\int^1_0 dx \ln\left[ m_{h}^2 x^2-x( m_{h}^2+M^2_1-M^2_2)+M_1^2-i\delta\right].\crn \label{b0function}\eea
 The $b^{(1)}_0$ can be found in a very simple form in the limit $p_i^2\rightarrow0$.  The $b_0^{(12)}$ is determined by
\be b_0^{(12)}=  -\sum_{k=1}^2\int^1_0 dx \ln(x-x_k), \label{B0f1}\ee
where $x_k,~(k=1,2)$ are solutions of the equation
 \be  x^2-\left(\frac{m_h^2-M_1^2+M_2^2}{m_h^2}\right)x+\frac{M_2^2-i\delta}{m_h^2}=0.\label{squeq}\ee
 The final expression of $b^{(12)}_0$ is
 \bea  b_0^{(12)}&=&\ln \frac{m_h^2-i\delta}{M_1^2-i\delta}+2 + \sum_{k=1}^2 x_k\ln\left(1-\frac{1}{x_k}\right).   \label{b012f1}\eea

The $B^{i}_1, ~B^{(12)}_i$ are calculated through the $B_0$ and $A_0$ functions, namely
\bea
B^{(i)}_1&=&\frac{(-1)^{i-1}}{2 m^2_i}\left[ A_0(M_i)-A_0(M_0)+B^{(i)}_0(M_0^2-M_i^2+m_i^2)\right], \crn
B^{(12)}_i&=&\frac{1}{2 m^2_h}\left[ A_0(M_1)-A_0(M_2)+B^{(12)}_0\left(M_2^2-M_1^2+(-1)^{i-1}m_h^2\right)\right] .\crn \label{bifunction}\eea
The $C_i$ functions can be found through the equation
 \bea && \left(
       \begin{array}{cc}
         2m^2_1 & m^2_h-m_1^2-m_2^2 \\
         m^2_h-m_1^2-m_2^2 & 2 m_2^2 \\
       \end{array}
     \right) \left(
               \begin{array}{c}
                 C_1 \\
                 C_2 \\
               \end{array}
             \right)\crn
             &=&\left(
                       \begin{array}{c}
                         B^{(12)}_0-B^{(2)}_0+ (M_0^2-M_1^2+m_1^2)C_0 \\
                         -\left[B^{(12)}_0 -B^{(1)}_0+ (M_0^2-M_2^2+m_2^2)C_0\right] \\
                       \end{array}
                     \right). \label{Cif2}
 \eea
 The $C_0$ function was generally calculated in \cite{Hooft}, a more explicit explanation was given in \cite{Ninhthes}. In the limit $p_1^2,p_2^2\rightarrow 0$,
 we get the following expression  \bea C_0 &=& - \int^{1}_{0}dx\int^{1-x}_0\frac{dy}{(1-x-y)M_0^2+xM_1^2+yM_2^2-xy m_h^2-i\delta}\crn
 &=&\frac{1}{m_h^2} \int^{1}_{0}\frac{dx}{x-x_0}\crn
 &\times&\left[\ln\frac{m_h^2-i\delta'}{M_1^2-M_0^2-i\delta'}+\ln(x-x_1)+\ln(x-x_2)- \ln(x-x_3)\right]\crn
 &=&\frac{1}{m_h^2} \ln\frac{m_h^2-i\delta'}{M_1^2-M_0^2-i\delta'}\times \ln\left(1-\frac{1}{x_0}\right)
 \crn
 &+&\frac{1}{m_h^2} \int^{1}_{0}\frac{dx}{x-x_0}\left[\ln(x-x_1)+\ln(x-x_2)- \ln(x-x_3)\right], \label{C0f1}\eea
 where both  $\delta $ and $\delta'$ are positive and extremely small,  $x_0$ and $x_3$ are defined as
 \be x_0=\frac{M_2^2-M_0^2}{m_h^2},\hs x_3=\frac{-M_0^2+i\delta}{M_1^2-M_0^2}, \label{x13}\ee
 and $x_1,x_2$ are solutions of the equation (\ref{squeq}). The limit of $p_1^2,p_2^2=0$  will be used in our work, even  when the loops contain active neutrinos with masses extremely smaller than  these quantities, because of the appearance of heavy virtual particles.  The explanation is as follows. The denominator in the first line of (\ref{C0f1}) has the general form of $D=(1-x-y)M_0^2+xM_1^2+yM_2^2-xy m_h^2-i\delta -(1-x-y)\left[x m_1^2+ym_2^2\right]$. Our calculation relates to the two following cases:
 \begin{itemize}
         \item Only $M_0$ is the mass of the active neutrino, $ M_0\ll M_1,M_2$.  We have $D=(1-x-y)M_0^2+xM_1^2\left[1-(1-x-y)m_1^2/M_1^2 \right] +yM_2^2\left[1-(1-x-y)m_2^2/M_2^2 \right] -xy m_h^2-i\delta \simeq (1-x-y)M_0^2+xM_1^2+yM_2^2-xy m_h^2-i\delta$.
         \item $M_1=M_2$ is the mass of the neutrino: $M_1=M_2\ll M_0$.  Then we have $D=(1-x-y)M_0^2\left[1-(xm_1^2+ym_2^2)/M_0^2\right]+xM_1^2+yM_2^2-xy m_h^2-i\delta\simeq (1-x-y)M_0^2+xM_1^2+yM_2^2-xy m_h^2-i\delta$.
       \end{itemize}
We use the following result given in \cite{Hooft}
\bea R(x_0,x_i)&\equiv&\int_0^1\frac{dx}{x-x_0}\left[\ln(x-x_i)-\ln(x_0-x_i) \right]\crn&=& Li_2(\frac{x_0}{x_0-x_i})- Li_2(\frac{x_0-1}{x_0-x_i}),\label{integ1}\eea
where $i=1,2,3$ and $Li_2(z)$ is the di-logarithm defined by
$$ Li_2(z) \equiv\int_0^1- \frac{dt}{t}\ln(1-tz).$$  We also use the real values of $x_0$ to give the result $ \eta(-x_i,\frac{1}{x_0-x_i})\ln\frac{x_0}{x_0-x_i}= \eta(1-x_i,\frac{1}{x_0-x_i})\ln\frac{x_0-1}{x_0-x_i}=0$ for any complex $x_i$.
Now we introduce the function
\bea R_0(x_0,x_i) \equiv Li_2(\frac{x_0}{x_0-x_i})- Li_2(\frac{x_0-1}{x_0-x_i}), \label{r0function}\eea
leading to
\be \int_0^1\frac{dx\ln(x-x_i)}{x-x_0}=R_0(x_0,x_i)+ \ln\left(1-\frac{1}{x_0}\right)\ln(x_0-x_i).\label{cint} \ee
Using the following equalities
 $$ \ln (AB-i\delta)=\ln(A-i\delta')+ \ln(B-i\delta/A)$$ with any real $A,B$,   $\delta,\delta'$ positive real and extremely small;  and
 $$x_1x_2=\frac{m^2_h-M_1^2+M_2^2}{m_h^2}, \hs x_1 x_2=\frac{M_2^2-i\delta}{m^2_h},$$
  we can prove that
  $$  \ln\frac{m_h^2-i\delta'}{M_1^2-M_0^2-i\delta'}+\ln(x_0-x_1)+\ln(x_0-x_2)-\ln(x_0-x_3)=0. $$
  This results the very simple expression of $C_0$ function
 \be  C_0=\frac{1}{m_h^2}\left[R_0(x_0,x_1)+ R_0(x_0,x_2)-R_0(x_0,x_3)\right] , \label{C0fomula1}\ee
 where $x_{1,2}$ are solutions  of  the equation (\ref{squeq}), and $x_{0,3}$ are given in (\ref{x13}).  This result is consistent with that discussed on \cite{bardin}.

 For  simplicity  in calculation we will also use  other approximations of PV-functions where $p_1^2,p_2^2\rightarrow 0$, namely
\bea a_0(M) &=& M^2\left(1+\ln\frac{m_h^2-i\delta}{M^2-i\delta}\right) , \hs
b^{(i)}_0 =1-\ln\frac{M_i^2}{m_h^2}+\frac{M_0^2}{M_0^2-M_i^2}\ln\frac{M_i^2}{M_0^2},\crn
 b^{(1)}_1  &=& -\frac{1}{2}\ln\frac{M_1^2}{m_h^2}-\frac{M_0^4}{2(M_0^2-M_1^2)^2}\ln\frac{M_0^2}{M_1^2} +\frac{(M_0^2-M^2_1)(3 M_0^2-M_1^2)}{4(M_0^2-M_1^2)^2}, \crn
  b^{(2)}_1  &=& \frac{1}{2}\ln\frac{M_2^2}{m_h^2}+\frac{M_0^4}{2(M_0^2-M_2^2)^2}\ln\frac{M_0^2}{M_2^2} -\frac{(M_0^2-M^2_2)(3 M_0^2-M_2^2)}{4(M_0^2-M_2^2)^2}, \crn
b^{(12)}_0&=&\ln \frac{m_h^2-i\delta}{M_1^2-i\delta}+2 + \sum_{k=1}^2 x_k\ln\left(1-\frac{1}{x_k}\right),\nonumber \label{B0f2}\eea
where $x_k$ is the two solutions of the equation (\ref{squeq}),
\bea
b^{(12)}_i &=& \frac{1}{2 m^2_h}\left[ M_1^2\left(1+\ln\frac{m_h^2}{M_1^2}\right)- M_2^2\left(1+\ln\frac{m_h^2}{M_2^2}\right)\right]\crn
&&+  \frac{b^{(12)}_0}{2 m^2_h}\left[M_2^2-M_1^2+(-1)^{i-1}m_h^2\right],\crn
 C_1 &=& \frac{1}{m_h^2}   \left[b^{(1)}_0 -b_0^{(12)}+(M_2^2-M_0^2)C_0\right],\crn
  C_2 &=&  -\frac{1}{m_h^2}   \left[b^{(2)}_0 -b_0^{(12)}+(M_1^2-M_0^2)C_0\right]. \nn\eea

\section{\label{Camlitude}Calculations the one loop contributions}
 In the first part of this section we will calculate in details the contributions of particular contributions of diagrams shown in the figure \ref{nDiagram} which involve  with exotic neutral lepton $N_a$, $a=1,2,3$. From this we can derive the general functions expressing the contributions of particular diagrams.

 \subsection{Amplitudes}
 It is needed to remind that the amplitude will be expressed in terms of the PV-functions, so the integral will be written as
  $$ \int \frac{d^4 k}{(2\pi)^4} \rightarrow  \frac{i}{16\pi^2}\times \frac{(2\pi\mu)^{4-D}}{i\pi^2} \int d^4 k,$$ where $\mu$ is a parameter with dimension of mass. This step will be omitted in the below calculation,  the final results are simply corrected  by adding  the factor $i/16\pi^2$. As an example in the calculation of contribution from the first diagram, we will point out a class of divergences that automatically vanish by the GIM mechanism. More explicitly for  any terms which do not depend on the masses of virtual leptons, they will vanish because of  the appearance of the factor $\sum_{a}V^{L}_{1a}V^{L*}_{2a}=0$.

 The contribution from diagram \ref{nDiagram}a) is: {\small}
{\footnotesize\bea i \mathcal{M}_{(a)}^{FVV}&=&\sum_{a}\int \fr{d^4k}{(2\pi)^4}\times\bar{u}_1\fr{ig}{\sqrt{2}} V_{1a}\ga^\mu P_L\fr{1(/\!\!\!k+m_a)}{D_0}\fr{ig}{\sqrt{2}} V^*_{2a}\ga^\nu P_Lv_2\crn
&&\times \left[\fr{ig m_V}{\sqrt{2}} \left(-c_{\alpha}s_{\theta}+\sqrt{2}s_{\alpha}c_{\theta} \right) \right] \fr{-i}{D_1} \crn
&&\times\left[g_{\mu\al}-\fr{(k-p_1)_\mu(k-p_1)_\al}{m_V^2} \right]\fr{-i}{D_2}\left[g_{\nu\beta}-\fr{(k+p_2)_\nu(k+p_2)_\bet}{m_V^2} \right]\crn
&=& \sum_{a} V_{1a}V^*_{2a}(-1) \fr{g^3 m_V}{2\sqrt{2}} \left(-c_{\alpha}s_{\theta}+\sqrt{2}s_{\alpha}c_{\theta} \right) \times \int\fr{d^4k}{(2\pi)^4}  \fr{\bar{u}_1\ga^{\mu}/\!\!\!k\ga^{\nu}P_Lv_2}
{D_0D_1D_2} \crn
&&\times  g^{\al\bet} \left[ g_{\mu\al}-\fr{(k-p_1)_{\mu}(k-p_1)_{\al}}{m^2_V}\right]\left[ g_{\bet\nu}-\fr{(k+p_2)_{\nu}(k+p_1)_{\bet}}{m^2_V}\right]\crn
&\equiv & \sum_{a} V_{1a}V^*_{2a}(-1) \fr{g^3 m_V}{2\sqrt{2}} \left(-c_{\alpha}s_{\theta}+\sqrt{2}s_{\alpha}c_{\theta} \right)  \left[P_1+P_2+P_3\right], \label{Di1a}\eea}
where
{\footnotesize\bea
P_1 &=& \int\fr{d^4k}{(2\pi)^4}\fr{\bar{u}_1\ga^\mu /\!\!\!k \ga^\nu P_L v_2}{D_0D_1D_2}g_{\mu\nu} = \int \fr{d^4k}{(2\pi)^4}\fr{(2-d)\bar{u}_1/\!\!\!k P_L v_2}{D_0D_1D_2} \crn
&=&\bar{u}_1 P_L v_2\times m_1(-2C_1) +\bar{u}_1 P_L v_2\times m_2(2C_2),\label{Di1a1}
\eea}
We can see that $P_1$ does not contain  any divergent terms. The formula of $P_2$ is
{\footnotesize\bea
P_2 &=&
\fr{-1}{m^2_V}\int\fr{d^4k}{(2\pi)^4}\fr{\bar{u}_1\ga^\mu /\!\!\!k \ga^\nu P_L v_2}{D_0D_1D_2}\left[(k+p_2)_\mu(k+p_2)_\nu+(k-p_1)_\mu(k-p_1)_\nu \right]\crn
&=&\fr{-1}{m^2_V}\int\fr{d^4k}{(2\pi)^4}\left[\fr{\bar{u}_1 (D_0+m^2_a)(/\!\!\!k+2/\!\!\!p_2)P_Lv_2+\bar{u}_1/\!\!\!p_2/\!\!\!k/\!\!\!p_2 P_Lv_2}{D_0D_1D_2}\right. \crn
&&+\left. \fr{\bar{u}_1(D_0+m^2_a)(/\!\!\!k-2/\!\!\!p_1)P_Lv_2+\bar{u}_1/\!\!\!p_1/\!\!\!k/\!\!\!p_1 P_Lv_2}{D_0D_1D_2} \right]\crn
&=&\fr{-1}{m^2_V}\left\{\fr{}{}\bar{u}_1P_Lv_2\times m_1 \left[\fr{}{}2B^{(12)}_1(m_V)-2B^{(12)}_0(m_V)\right.\right.\crn
&&\left.\left. -2m_a^2C_0 +(2m^2_a+m^2_1-m_2^2)C_1+(m^2_{H_0}-m_1^2-m_2^2)C_2 \dfrac{}{}\right]\right.\crn
&&\left.+\bar{u}_1P_Rv_2\times m_2 \left[\fr{}{}-2B^{(12)}_2(m_V)-2B^{(12)}_0(m_V)\right.\right.\crn
&&\left.\left. -2m_a^2C_0-(2m^2_a-m^2_1+m^2_2)C_2-(m^2_{H_0}-m_1^2-m^2_2)C_1 \fr{}{}\right]\right\}. \label{Di1a2}
\eea{\footnotesize
We can see that the terms like $ B^{(12)}_1(m_V),~B^{(12)}_1(m_V)$ and $B^{(12)}_0(m_V) $ do contain divergences but theydo  not depend on $m_a$ in the loop.   Hence  these terms will exactly cancel by  the GIM mechanism.  All of the other are finite.

The contribution from $P_3$ is
{\footnotesize\bea
P_3 &=& \fr{1}{m^4_V}\int\fr{d^4k}{(2\pi)^4}\times\fr{\bar{u}_1\ga^\mu/\!\!\!k\ga^\nu P_Lv_2}{D_0D_1D_2}\left[(k-p_1).(k+p_2)(k-p_1)_\mu(k+p_2)_\nu \right] \crn &=&\fr{1}{2m^4_V}\int\fr{d^4k}{(2\pi)^4}\times \left[\fr{\bar{u}_1[D_1+D_2+2m_V^2-m^2_{H_0}](D_0+m_a^2)(/\!\!\!k+/\!\!\!p_2-/\!\!\!p_1)P_Lv_2}{D_0D_1D_2}\right.\crn
&&\left. +m_1m_2\fr{\bar{u}_1[D_1+D_2+2m_V^2-m^2_{H_0}]/\!\!\!k P_Lv_2}{D_0D_1D_2}\right]\crn
&=&\fr{1}{2m^4_V}\left\{\fr{}{}\bar{u}_1P_Lv_2\times m_1\left[-A_0(m_V)+(2m_V^2-m^2_{H_0})\left( B^{(12)}_1(m_V)-B^{(12)}_0(m_V)\right)\right.\right.\crn
&&\left.\left. -m_2^2B^{(2)}_1+  \mathbf{ m^2_a\left(B^{(1)}_1-B^{(1)}_0-B^{(2)}_0\right)}\right.\right.\crn
 &&\left.\left.+(2m_V^2-m^2_{H_0})\left(\fr{}{} m^2_a(C_1-C_0)-m_2^2C_2\right)\right]\right.\crn
&&\left.+ \bar{u}_1P_Rv_2\times m_2\left[-A_0(m_V)+(2m_V^2-m^2_{H_0})\left(-B^{(12)}_2(m_V)-B^{(12)}_0(m_V)\right)\right.\right.\crn
&&\left.\left.+m_1^2B^{(1)}_1  + \mathbf{m_a^2\left(-B^{(1)}_0-B^{(2)}_0-B^{(2)}_1\right)}\right.\right.\crn&& \left.\left. +(2m_V^2-m^2_{H_0})\left(\fr{}{} m_1^2C_1-m^2_a(C_0+C_2)\right)\right]\right\}.  \label{Di1a3}
\eea}
Again all terms in the first and third lines do not contribute to the amplitude. But  the four terms  $m_2^2B^{(2)}_1,$  $ m_1^2B^{(1)}_1$, $ m^2_a\left(B^{(1)}_1-B^{(1)}_0-B^{(2)}_0\right)$ and $m_a^2\left(-B^{(1)}_0-B^{(2)}_0-B^{(2)}_1\right)$ do contain divergences. The  first two terms have divergent parts  having the corresponding forms of $ (-m_2^2\Delta_{\epsilon})$  and $m_1^2\Delta_{\epsilon}$, which do not depend on the masses $m_a$ of the virtual leptons.  Hence  they also vanish by the GIM mechanism.  The finite parts of these terms still contribute to the amplitude.  The remain two  terms  include the most dangerous divergent parts. They  have factors $m_a^2$ which can not cancel by the GIM mechanism. We remark them by the bold and  will prove later that  they finally vanish after summing all diagrams. From now on we can exclude all terms that do not depend on the masses of virtual leptons.

 Based on definition $\mathcal{M}= -\left(E^{FVV}_L \overline{u_1}P_L v_2 + E^{FVV}_R   \overline{u_1}P_R v_2\right)$,  the expression of the total contribution from the diagram  \ref{nDiagram}a) is simply
{\footnotesize  \be \mathcal{M}_{(a)}^{FVV} =  \fr{-g^3 }{32 \pi^2\sqrt{2}} \left(-c_{\alpha}s_{\theta}+\sqrt{2}s_{\alpha}c_{\theta} \right)
    \sum_{a} V_{1a}V^*_{2a} \left[ \left(\overline{u_1} P_L v_2\right)   E^{FVV}_L  + \left( \overline{u_1} P_R v_2\right)   E^{FVV}_R  \right],\label{Di1ap} \ee}
where $E^{FVV}_{L,R}$ is defined in (\ref{EfvvL}) and (\ref{EfvvR}).
 Here we have added a factor of $\frac{i}{16\pi^2}$. All  terms being independent on $m_a$ will cancel by the factor  $\sum_{a} V_{1a}V^*_{2a}$.  If we assume all other divergences cancel among themselves  after summing all of the diagrams, the analytic formulas of $E^{FVV}_L$ and $E^{FVV}_R$  can be written in  terms of the finite parts of  PV-functions, i.e $b^{(i)}_{0}, b^{(12)}_0, b^{i}_1,\; b^{(12)}_i$ and $C_{0,1,2}$.  The following calculation for the remain diagrams will be done the same as what we have done above. We trace the divergence of each diagram in the bold text.

The contribution from diagram \ref{nDiagram}b) is:
{\footnotesize\bea
i\mathcal{M}^{FVH}_{(b)}&=& \sum_{a}\int\fr{d^4k}{(2\pi)^4} \crn
&\times&\bar{u}_1\fr{ig}{\sqrt{2}}V_{1a}\ga^\mu P_L\fr{i(/\!\!\!k+m_a)}{D_0}(-i\sqrt{2}V_{2a}^*)\left(\fr{m_2}{v_1}a_1P_R+ \fr{m_a}{v_3}a_3P_L\right)v_2
\crn
&&\times\fr{ig}{2\sqrt{2}}(-k-2p_2-p_1)^\al\fr{i}{D_2}\fr{-i}{D_1}\left[g_{\mu\al} -\fr{(k-p_1)_\mu(k-p_1)_\al}{m^2_V}\right] \crn
 &=&
\sum_{a}V_{1a}V^*_{2a}\fr{g^2}{2\sqrt{2}}(c_\al c_\theta+\sqrt{2}s_\al s_\theta)\int\crn
 &&\times \fr{d^4k}{(2\pi)^4}\times \fr{\fr{m_2}{v_1}a_1\bar{u}_1\ga^\mu /\!\!\!k P_Rv_2+\fr{m^2_a}{v_2}a_2\bar{u}_1\ga^\mu P_Lv_2}{D_0D_1D_2}\crn
&&\times\left[(k+2p_2+p_1)_\mu-\fr{(k+2p_2+p_1).(k-p_1)(k-p_1)_\mu}{m_V^2} \right]\crn
&=&\sum_{a}V_{1a}V^*_{2a} \left[\fr{g^2}{2\sqrt{2}}(c_\al c_\theta+\sqrt{2}s_\al s_\theta)\right] \crn
&&\times \left\{\fr{}{}\bar{u}_1P_Lv_2\times \left[\fr{}{}\fb{$\bf\fr{-m_1}{m^2_V}\fr{m^2_a}{v_3}a_3\left(B_1^{(1)} -B^{(1)}_0 \right)$}\right.\right.\crn
&&\left.\left. +\fr{m^2_a}{v_3}a_3\times m_1\left(C_0+C_1 +\fr{(m^2_{H_A}-m^2_{H_0})}{m_V^2}(C_0-C_1)\right)\right.\right.\crn
&&\left.\left. +\fr{m_2}{v_1}a_1\times m_1m_2\left(2C_1-C_2-\fr{m^2_{H_A}-m^2_{H_0}}{m_V^2}C_2\right)\right]\right.\crn
&&\left.+\bar{u}_1P_Rv_2\times \left[\dfrac{-1}{m_V^2}\fr{m_2}{v_1}a_1\left( A_0(m_V)+(m^2_{H_A}-m^2_{H_0})B_0^{(12)}\right)\right.\right.\crn
&&\left.\left. +\fr{m_2}{v_1}a_1B_0^{(12)}  +\fr{m_1^2}{m_V^2}\fr{m_2}{v_1}a_1B_1^{(1)} -\fb{$\bf\dfrac{m_a^2}{m_V^2}\dfrac{m_2}{v_1}a_1B^{(1)}_0(m_a,m_V)$}\right.\right.\crn
&&\left.\left. +\fr{m_2}{v_1}a_1\left(m_a^2C_0-m^2_1C_1+2m_2^2C_2+2(m^2_{H_0}-m_2^2) C_1 \right.\right. \right. \crn&&- \left.\left.\left.  \fr{(m^2_{H_A}-m^2_{H_0})}{m_V^2}\left(m_a^2C_0-m_1^2C_1\right)\right)\right.\right.\crn
&&\left.\left. +\fr{m_a^2}{v_3}a_3\times m_2\left(-2C_0-C_2+\fr{(m^2_{H_A}-m^2_{H_0})}{m_V^2}C_2\right)\fr{}{}\right]\right\}.\label{Di1b}
\eea}
The contribution to the total amplitude is
\be \mathcal{M}_{(b)}^{FVH} =  \fr{g^2 }{32 \pi^2\sqrt{2}} \left(c_{\alpha}c_{\theta}+\sqrt{2}s_{\alpha}s_{\theta} \right)   \sum_{a} V_{1a}V^*_{2a} \left[ \left(\overline{u_1} P_L v_2\right)   E^{FVH}_L  + \left( \overline{u_1} P_R v_2\right)   E^{FVH}_R  \right],\label{Dibp} \ee
The contribution from diagram \ref{nDiagram}c)  is:
{\footnotesize\bea
i\mathcal{M}^{FHV}_{(c)}&=&\sum_{a}\int\fr{d^4k}{(2\pi)^4}\times \bar{u}_1(-i\sqrt{2}V_{1a})\left(\fr{m_1}{v_1}a_1P_L+ \fr{m_a}{v_3}a_3P_R\right) \crn
&& \times \fr{i(/\!\!\!k+m_a)}{D_0}\fr{ig}{\sqrt{2}}V^*_{2a}\ga^\mu P_Lv_2
\times\fr{ig}{2\sqrt{2}}(c_\al c_\theta+\sqrt{2}s_\al s_\theta)(-k+p_2+2p_1)^\al
\crn&\times&\fr{i}{D_1}\fr{-i}{D_2} \times \left[g_{\mu\al} -\fr{(k+p_2)_\mu(k+p_2)_\al}{m^2_V}\right]\crn
&=&\sum_{a}V_{1a}V^*_{2a} \fr{g^2}{2\sqrt{2}}(c_\al c_\theta+\sqrt{2}s_\al s_\theta)\int \fr{d^4k}{(2\pi)^4}\crn
&&\times
\left[\fr{m_1}{v_1}a_1\fr{\bar{u}_1\ga^\mu /\!\!\!k P_Lv_2}{D_0D_1D_2}+\fr{m^2_a}{v_3}a_3\fr{\bar{u}_1\ga^\mu P_Lv_2}{D_0D_1D_2}\right]\crn
&&\times \left[(k-p_2-2p_1)_\mu -\fr{(k-p_2-2p_1).(k+p_2)(k+p_2)_\mu}{m_V^2} \right]\crn
&=& \sum_{a}V_{1a}V^*_{2a} \left[\fr{g^2}{2\sqrt{2}}(c_\al c_\theta+\sqrt{2}s_\al s_\theta)\right] V_{1a}V^*_{2a}\crn
&&\times\left\{\fr{}{}\bar{u}_1P_Lv_2\times \left[\fr{-1}{m_V^2}\fr{m_1}{v_1}a_1\left(A_0(m_V)+(m^2_{H_A}-m^2_{H_0})B_0^{(12)}\right)\right.\right.\crn
&&\left.\left. +\fr{m_1}{v_1}a_1B_0^{(12)}(m_V,m_{H_A})-\fr{m_2^2}{m_V^2}\fr{m_1}{v_1}a_1 B^{(2)}_{1}(m_a,m_V)\right.\right. \crn
&-&\left.\left. \fb{$\bf\dfrac{m_a^2}{m_V^2}\dfrac{m_1}{v_1}a_1B_0^{(2)}(m_a,m_V)$}\right.\right.\crn
&&\left.\left. +\fr{m_1}{v_1}a_1\left(\fr{}{}m_a^2C_0-2m_1^2C_1+m_2^2C_2-2(m^2_{H_0}-m_1^2)C_2\right.\right.\right.\crn
&&\left.\left.\left. -\fr{(m^2_{H_A}-m^2_{H_0})}{m_V^2}(m_2^2C_2+m_a^2C_0) \right)\right.\right.\crn
&&\left.\left. +m_1\fr{m_a^2}{v_3}a_3\left(-2C_0+C_1-\fr{(m^2_{H_A}-m^2_{H_0})}{m_V^2}C_1 \right)\right]\right.\crn
&&\left. +\bar{u}_1P_Rv_2\left[\fr{}{}\fb{$\bf\dfrac{m_2}{m_V^2}\dfrac{m_a^2}{v_3}a_3\left( B_1^{(2)}+B_0^{(2)}\right)$} \right.\right.\crn &&\left.\left.+m_1m_2\fr{m_1}{v_1}a_1\left(C_1-2C_2+ \fr{(m^2_{H_A}-m^2_{H_0})}{m_V^2}C_1\right)\right.\right.\crn
&&\left.\left. +m_2\fr{m_a^2}{v_3}a_3\left(C_0-C_2+\fr{(m^2_{H_A}-m^2_{H_0})}{m_V^2}(C_0+C_2)\right)\fr{}{}\right]\right\}. \label{Di1c}
\eea}
The contribution to the total amplitude is
\be \mathcal{M}_{(c)}^{FHV} =  \fr{g^2 }{32 \pi^2\sqrt{2}}\left(c_{\alpha}c_{\theta}+\sqrt{2}s_{\alpha}s_{\theta} \right)
\sum_{a} V_{1a}V^*_{2a} \left[ \left(\overline{u_1} P_L v_2\right)   E^{FHV}_L  + \left( \overline{u_1} P_R v_2\right)   E^{FHV}_R  \right]. \label{Dibp} \ee
 The contribution from diagram \ref{nDiagram}d)  is:
{\footnotesize\bea
i\mathcal{M}^{FHH}_{(d)}&=&\sum_{a}\int\fr{d^4k}{(2\pi)^4}\times (-iv_3\lambda_{h^0H_1H_1})\fr{i}{D_1}\fr{i}{D_2}\times \bar{u}_1(-i\sqrt{2}V_{1a}) \crn
&\times& \left(\fr{m_1}{v_1}a_1P_L+ \fr{m_a}{v_3}a_3P_R\right)\fr{i(/\!\!\!k+m_a)}{D_0}(-i\sqrt{2}V^*_{2a})\left(\fr{m_2}{v_1}a_1P_R+ \fr{m_a}{v_3}a_3P_L\right)v_2
\crn
&=&\sum_{a}v_3\lambda_{h^0H_1H_1} V_{1a}V^*_{2a}\int\dfrac{d^4k}{(2\pi)^4} \crn
&\times&\fr{\bar{u}_1\left(\fr{m_1}{v_1}a_1P_L+ \fr{m_a}{v_3}a_3P_R\right)(/\!\!\!k+m_a)\left(\fr{m_2}{v_1}a_1P_R+ \fr{m_a}{v_3}a_3P_L\right)v_2}{D_0D_1D_2}\crn
&=&\sum_{a}v_3\lambda_{h^0H_1H_1} V_{1a}V^*_{2a}\int\fr{d^4k}{(2\pi)^4} \crn
&\times& \left[\dfrac{m_1m_2}{v^2_1}a_1^2\dfrac{\bar{u}_1/\!\!\!k P_Rv_2}{D_0D_1D_2}+\dfrac{m_1m^2_a}{v_1v_3}a_1a_3\fr{\bar{u}_1 P_Lv_2}{D_0D_1D_2}\right.\crn
&&\left. +\fr{m^2_a}{v^2_3}a_3^2\fr{\bar{u}_1/\!\!\!k P_Lv_2}{D_0D_1D_2}+\fr{m_2m^2_a}{v_1v_3}a_1a_3\fr{\bar{u}_1 P_Rv_2}{D_0D_1D_2}\right]\crn
&=&\sum_{a}v_3 \lambda_{h^0H_1H_1} V_{1a}V^*_{2a}\crn
&\times&\left\{\bar{u}_1P_Lv_2\times m_1\left[\fr{m_a^2}{v_1v_3}a_1a_3C_0-\fr{m_2^2}{v^2_1}a^2_1C_2+\fr{m_a^2}{v^2_3}a^2_3C_1\right]\right.\crn
&&\left.+\bar{u}_1P_Rv_2\times m_2\left[\fr{m_a^2}{v_1v_3}a_1a_3C_0+\fr{m_1^2}{v^2_1}a^2_1C_1-\fr{m_a^2}{v^2_3}a^2_3C_2\right]\fr{}{}\right\}\label{Did}
\eea}
with $\lambda_{h^0H_1H_1}$ shown in the table \ref{smh0coupl}. With the notations of $E^{FHH}_L$ and $E^{FHH}_R$  defined in  (\ref{EfhhL}) and (\ref{EfhhR}), the contribution to the  amplitude is
{\footnotesize\be \mathcal{M}^{FHH}_{(d)} = \fr{1}{64 \pi^2\sqrt{2}}\times (4\sqrt{2}\lambda_{h^0H_1H_1}) \sum_{a}V_{1a}V^*_{2a} \left[\left(\overline{u_1} P_L v_2\right)   E^{FHH}_L+   \left( \overline{u_1} P_R v_2\right)  E^{FHH}_R \right]. \label{Did}\ee}
 The contribution from diagram \ref{nDiagram}e) is:
{\footnotesize\bea
i\mathcal{M}^{VFF}_{(e)}&=&\sum_{a} \int\fr{d^4k}{(2\pi)^4}\times \bar{u}_1 \fr{ig}{\sqrt{2}}V_{1a}\ga^\mu P_L\fr{i(-/\!\!\!k+/\!\!\!p_1+ m_a)}{D_1}\left(\fr{-igm_a}{2m_V}\fr{s_\al}{c_\theta}\right) \crn
&& \times\fr{i(-/\!\!\!k-/\!\!\!p_2+ m_a)}{D_2} \fr{ig}{\sqrt{2}}V^*_{2a}\ga^\nu P_Lv_2 \fr{-i}{D_0}\left[g_{\mu\nu}-\fr{k_\mu k_\nu}{m_V^2} \right] \crn
&=&\sum_{a}-\fr{g^3m_a}{4m_V}\fr{s_\al}{c_\theta}V_{1a}V^*_{2a}\int\fr{d^4k}{(2\pi)^4} \left[\fr{(2-d)m_a\bar{u}_1(-2/\!\!\!k+/\!\!\!p_1-/\!\!\!p_2)P_Lv_2}{D_0D_1D_2}\right.\crn
&&\left.-\frac{m_a}{m^2_V}\fr{\bar{u}_1/\!\!\!k(-2/\!\!\!k+/\!\!\!p_1-/\!\!\!p_2)/\!\!\!kP_Lv_2}{D_0D_1D_2}\fr{}{}\right]\crn
&=&\sum_{a}\left[-\fr{g^3m_a}{4m_V}\fr{s_\al}{c_\theta}\right] V_{1a}V^*_{2a}\left\{\fr{}{}\bar{u}_1P_Lv_2\times m_1m_a\left[\fr{}{} \fb{$\bf\fr{1}{m_V^2}\left(\dfrac{}{}B^{(12)}_{0}+B^{(1)}_1\right)$}\right.\right.\crn
&&\left.\left.-\fr{1}{m_V^2}\left(-m^2_VC_0 +(m_1^2+m_2^2-2m_a^2)C_1\fr{}{}\right)+(2-d)(C_0-2C_1)\right]\right.\crn
&&\left. +\bar{u}_1P_Rv_2\times m_2m_a\left[\fr{}{}\fb{$\bf\fr{1}{m_V^2}\left(\dfrac{}{}B^{(12)}_{0}-B^{(2)}_1 \right)$}+(2-d)(C_0+2C_2) \right.\right.\crn
&&\left.\left. -\fr{1}{m_V^2}\left(-m^2_VC_0-(m_1^2+m_2^2-2m_a^2)C_2\fr{}{}\right)\right]\right\}. \label{Di1e}
\eea}
The final result is written as
\be  \mathcal{M}^{VFF}_{(e)} = \left[-\fr{1}{64 \pi^2 \sqrt{2}}\times \fr{g^3s_\al\sqrt{2}}{c_\theta}\right]  \sum_{a}V_{1a}V^*_{2a}
\left[ \left(\overline{u_1} P_L v_2\right)E^{VFF}_L + \left( \overline{u_1} P_R v_2\right)  E^{VFF}_R \right], \label{Diep} \ee
where $E^{VFF}_{L,R}$ are defined in (\ref{EvffL}) and (\ref{EvffR}).

The contribution from diagram \ref{nDiagram}f) is
{\footnotesize\bea
i\mathcal{M}^{HFF}_{(f)}&=&\sum_{a}\int\fr{d^4k}{(2\pi)^4} \times \bar{u}_1(-i\sqrt{2}V_{1a})\left(\fr{m_1}{v_1}a_1P_L+ \fr{m_a}{v_3}a_3P_R\right) \crn
&\times&\fr{i(-/\!\!\!k+/\!\!\!p_1+m_a)}{D_1}\left(\fr{-im_as_\al}{v_3}\right)\fr{i(-/\!\!\!k-/\!\!\!p_2+m_a)}{D_2} \crn
&&\times (-i\sqrt{2}V^*_{2a})\left(\fr{m_2}{v_1}a_1P_R+ \fr{m_a}{v_3}a_3P_L\right)v_2\times\fr{i}{D_0}\crn
&=&\sum_{a}V_{1a}V^*_{2a}\left[\fr{2m_as_\al}{v_3}\right]\int\fr{d^4k}{(2\pi)^4} \left[\fr{m_1m_a}{v_1v_3}a_1a_3\fr{\bar{u}_1(/\!\!\!k-/\!\!\!p_1)(/\!\!\!k+/\!\!\!p_2)P_Lv_2}{D_0D_1D_2}\right.\crn
&&\left. +\fr{m_2m_a}{v_1v_3}a_1a_3\fr{\bar{u}_1(/\!\!\!k-/\!\!\!p_1)(/\!\!\!k+/\!\!\!p_2)P_Rv_2}{D_0D_1D_2}\right.\crn
&&\left.+ m_a\fr{m_1m_2}{v_1^2}a^2_1\fr{\bar{u}_1(-2/\!\!\!k-/\!\!\!p_2+/\!\!\!p_1)P_Rv_2}{D_0D_1D_2}\right.\crn
&&\left. +\fr{m^3_a}{v_3^2}a^2_3\fr{\bar{u}_1(-2/\!\!\!k-/\!\!\!p_2+/\!\!\!p_1)P_Lv_2}{D_0D_1D_2} \right.\crn
&&\left.+\fr{m_1m^3_a}{v_1v_3}a_1a_3\fr{\bar{u}_1P_Lv_2}{D_0D_1D_2}+\fr{m_2m^3_a}{v_1v_3}a_1a_3\fr{\bar{u}_1P_Rv_2}{D_0D_1D_2} \right]\crn
&=&\sum_{a}V_{1a}V^*_{2a}\left[\fr{2m_as_\al}{v_3}\right] \left\{\fr{}{}\bar{u}_1P_Lv_2 \right.\crn
&&\left. \times m_1m_a \left[\fr{}{}\fb{$\bf\dfrac{a_1a_3}{v_1v_3}B^{(12)}_{0}$}+\fr{m_2^2}{v_1^2}a_1^2(2C_2+C_0) +\fr{m_a^2}{v_3^2}a_3^2(C_0-2C_1)\right.\right.\crn
&&\left.\left.+\fr{a_1a_3}{v_1v_3}\left(2m_2^2C_2-(m_1^2+m_2^2)C_1 +(m_a^2+m^2_{H_A}+m_2^2)C_0\fr{}{}\right)\right]\right.\crn
&&\left. +\bar{u}_1P_Rv_2m_2m_a \left[\fr{}{}\fb{$\bf\dfrac{a_1a_3}{v_1v_3}B^{(12)}_{0}$} +\dfrac{m_1^2}{v_1^2}a_1^2(C_0-2C_1)+\fr{m_a^2}{v_3^2}a_3^2(C_0+2C_2)\right.\right.\crn
&&\left.\left. +\fr{a_1a_3}{v_1v_3}\left(-2m_1^2C_1+(m_1^2+m_2^2)C_2+(m_a^2+m^2_{H_A}+m_1^2)C_0\fr{}{}\right)\right]\fr{}{}\right\}\label{Di1f}
\eea}
The final result is written as
{\footnotesize\be  i\mathcal{M}^{HFF}_{(f)} = \fr{1}{64\pi^2\sqrt{2}}\times (8s_{\alpha}\sqrt{2})  \sum_{a}V_{1a}V^*_{2a} \left[\left(\overline{u_1} P_L v_2\right)   E^{HFF}_L +  \left( \overline{u_1} P_R v_2\right)  E^{HFF}_R \right],  \label{Di1ep} \ee}
where $E^{HFF}_{L,R}$ are defined in (\ref{EhffL}) and (\ref{EhffR}).

The contribution from diagram \ref{nDiagram}g) is:
{\footnotesize\bea
i\mathcal{M}^{(FV)}_{(g)}&=& \sum_{a}\int\fr{d^4k}{(2\pi)^4} \times \bar{u}_1\left(\fr{ig}{\sqrt{2}} V_{1a}\ga^\mu P_L\right)\fr{i(/\!\!\!k+m_a)}{D_0}
\left(\fr{ig}{\sqrt{2}} V^*_{2a}\ga^\nu P_L\right)\crn
&&\times \fr{i(/\!\!\!p_1+m_2)}{p_1^2-m^2_2}\left(\fr{igm_2}{2\sqrt{2}m_V}\fr{c_\al}{s_\theta}\right)v_2
 \fr{-i}{D_1}\left[g_{\mu\nu}-\fr{(k-p_1)_\mu(k-p_1)_\nu}{m^2_V} \right]\crn
&=& \sum_{a}V_{1a}V^*_{2a}\fr{g^3}{4\sqrt{2}m_V}\fr{m_2}{(m_1^2-m_2^2)}\fr{c_\al}{s_\theta}\crn
&&\times\int\fr{d^4k}{(2\pi)^4}\left[\fr{(2-d)\bar{u}_1/\!\!\!k /\!\!\!p_1P_Rv_2 +(2-d)m_2\bar{u}_1 /\!\!\!k  P_Lv_2}{D_0D_1}\right.\crn
&&-\left.\fr{1}{m^2_V}\fr{\bar{u}_1(/\!\!\!k-/\!\!\!p_1)/\!\!\!k (/\!\!\!k-/\!\!\!p_1) /\!\!\!p_1P_Rv_2}{D_0D_1}-\fr{m_2}{m^2_V}\fr{\bar{u}_1(/\!\!\!k-/\!\!\!p_1)/\!\!\!k (/\!\!\!k-/\!\!\!p_1) P_Lv_2}{D_0D_1}\right]\crn
&=& \sum_{a}V_{1a}V^*_{2a} \left[\fr{g^3}{4\sqrt{2}m_V}\fr{m_2}{(m_1^2-m_2^2)}\fr{c_\al}{s_\theta}\right]\crn
&&\times\left\{\fr{}{}\bar{u}_1P_Lv_2\times m_1m_2\left[\fr{1}{m^2_V}A_0(m_V)-\fr{m_1^2}{m^2_V}B^{(1)}_1 \right.\right.\crn
&&\left.\left.+(2-d)B^{(1)}_1 -\fb{$\bf\fr{1}{m^2_V}\left(- 2m^2_aB^{(1)}_0 +m^2_aB^{(1)}_1 \fr{}{}\right)$}\fr{}{}\right]\right.\crn
&&\left. +\bar{u}_1P_Rv_2\times m^2_1 \left[\fr{1}{m^2_V}A_0(m_V)-\fr{m_1^2}{m^2_V}B^{(1)}_1+(2-d)B^{(1)}_1\right.\right.\crn    -&&\left.\left.\fb{$\bf\fr{1}{m_V^2}\left(-2m^2_aB^{(1)}_0 +m_a^2B^{(1)}_1 \right)$} \fr{}{}\right]\right\}\label{Di1g}
\eea}
The contribution from diagram \ref{nDiagram}h) is:
{\footnotesize\bea
i\mathcal{M}^{VF}_{(h)}&=& \sum_{a}\int\fr{d^4k}{(2\pi)^4} \times \bar{u}_1\left(\fr{igm_1}{2\sqrt{2}m_V}\fr{c_\al}{s_\theta}\right)\fr{i(-/\!\!\!p_2+m_1)}{p_2^2-m^2_1}\left(\fr{ig}{\sqrt{2}} V_{1a}\ga^\mu P_L\right)\crn
&&\times \fr{i(/\!\!\!k+m_a)}{D_0}\left(\fr{ig}{\sqrt{2}} V^*_{2a}\ga^\nu P_L\right) v_2 \times\fr{-i}{D_2}\left[g_{\mu\nu}-\fr{(k+p_2)_\mu(k+p_2)_\nu}{m^2_V} \right]\crn
&=& \sum_{a}\fr{g^3}{4\sqrt{2}m_V}\fr{m_1}{(m_2^2-m_1^2)} \fr{c_\al}{s_\theta}V_{1a}V^*_{2a}\crn
&&\times \int\fr{d^4k}{(2\pi)^4}\left[\fr{-(2-d)\bar{u}_1/\!\!\!p_2/\!\!\!k P_Lv_2 +(2-d)m_1\bar{u}_1 /\!\!\!k  P_Lv_2}{D_0D_2}\right.\crn
&&+\left.\fr{1}{m^2_V}\fr{\bar{u}_1 /\!\!\!p_2(/\!\!\!k+/\!\!\!p_2) /\!\!\!k (/\!\!\!k+/\!\!\!p_2)P_Lv_2}{D_0D_2}-\fr{m_1}{m^2_V}\fr{\bar{u}_1(/\!\!\!k+/\!\!\!p_2)/\!\!\!k (/\!\!\!k+/\!\!\!p_2) P_Lv_2}{D_0D_2}\right]\crn
&=& \sum_{a}\fr{g^3}{4\sqrt{2}m_V}\fr{m_1}{(m_2^2-m_1^2)}\fr{c_\al}{s_\theta}V_{1a} V^*_{2a}\crn
&&\times\int\fr{d^4k}{(2\pi)^4}\left[(2-d)\bar{u}_1\left(\fr{-/\!\!\!p_2/\!\!\!k}{D_0D_2}+\fr{m_1/\!\!\!k}{D_0D_2}\right)P_Lv_2\right.\crn
&&\left. +\fr{1}{m_V^2}\bar{u}_1\left(\fr{k^2/\!\!\!p_2/\!\!\!k+2k^2p^2_2+m^2_2/\!\!\!k/\!\!\!p_2}{D_0D_2}\right) P_Lv_2\right.\crn
&&\left.-\fr{m_1}{m_V^2}\bar{u}_1\left(\fr{k^2/\!\!\!k+2k^2/\!\!\!p_2+/\!\!\!p_2/\!\!\!k/\!\!\!p_2}{D_0D_2}\right)P_Lv_2\right]\crn
&=& \sum_{a}\left[\fr{g^3}{4\sqrt{2}m_V}\fr{m_1}{(m_2^2-m_1^2)}\fr{c_\al}{s_\theta}\right]V_{1a}V^*_{2a}\crn
&&\times\left\{\bar{u}_1P_Lv_2\times m^2_2\left[\fr{1}{m^2_V}A_0(m_V)+\fr{m_2^2}{m^2_V}B^{(2)}_1  -(2-d)B^{(2)}_1 \right.\right.\crn
&&\left.\left.  -\fb{$\bf\fr{1}{m_V^2}\left(-2m^2_aB^{(2)}_0 -m^2_aB^{(2)}_1 \right)$}\right]\right.\crn
&&\left.+\bar{u}_1P_Rv_2\times m_1m_2 \left[\fr{1}{m^2_V}\fr{}{}A_0(m_V)+\fr{m_2^2}{m^2_V}B^{(2)}_1 -(2-d)B^{(2)}_1 \right.\right.\crn
&&\left.\left.-\fb{$\bf\fr{1}{m_V^2}\left(-2m^2_aB^{(2)}_0 -m_a^2)B^{(2)}_1 \right)$}\fr{}{}\right]\right\}\label{Di1h}
\eea}
The total amplitude from the  two diagrams \ref{nDiagram}g) and \ref{nDiagram}h) is:
{\footnotesize\bea
i\mathcal{M}^{FV}_{(g+h)}&=& \sum_{a}\left[\fr{g^3}{4\sqrt{2}m_V}\fr{c_\al}{s_\theta}\right]V_{1a}V^*_{2a} \left\{\fr{}{}\bar{u}_1P_Lv_2\times\fr{m_1m_2^2}{(m_1^2-m_2^2)}\right.\crn
&&\left.\times\left[-2\left(B^{(1)}_1 +B^{(2)}_1 \right) -\fr{1}{m_V^2}\left(m_1^2B^{(1)}_1 +m_2^2B^{(1)}_1 \right)\right.\right.\crn
&&\left.\left.+\fb{$\bf\fr{m_a^2}{m_V^2}\left(2(B^{(1)}_0-B^{(2)}_0)-(B^{(1)}_1+B^{(2)}_1) \right)$}\fr{}{}\right]\right.\crn
&&\left. +\bar{u}_1P_Rv_2 \fr{m_1^2m_2}{m_1^2-m_2^2}\left[(2-d)\left(B^{(1)}_1 +B^{(2)}_1  \right) -\fr{1}{m_V^2}\left(m_1^2B^{(1)}_1+ m_2^2B^{(1)}_1 \right)\right.\right.\crn
&&\left.\left.+\fb{$\bf\fr{m_a^2}{m_V^2}\left(2\left(B^{(1)}_0-B^{(2)}_0\right)-\left(B^{(1)}_1+B^{(2)}_1 \right) \right)$}\fr{}{}\right]\right\}.
\eea}
We note that the divergence part in the above expression is zero. The final result is
\be \mathcal{M}^{FV}_{(g+h)} =\left[\fr{1}{64\pi^2\sqrt{2}}\times\fr{g^3c_\al}{s_\theta}\right] \sum_{a}V_{1a}V^*_{2a}
\left[ \left(\overline{u_1} P_L v_2\right) E^{FV}_L +\left( \overline{u_1} P_R v_2\right)  E^{FV}_R \right], \label{Di1ep} \ee
where $E^{FV}_{L,R}$ are defined in (\ref{DfvL}) and (\ref{DfvR}).

The contribution from the diagram \ref{nDiagram}i) is:
{\footnotesize\bea
i\mathcal{M}^{FH}_{(i)}&=&\sum_{a}\int\fr{d^4k}{(2\pi)^4} \times \bar{u}_1(-i\sqrt{2}V_{1a})\left(\fr{m_1}{v_1}a_1P_L+ \fr{m_a}{v_3}a_3P_R\right)\fr{i(/\!\!\!k+m_a)}{D_0}\crn
&&\times (-i\sqrt{2}V^*_{2a}) \left(\fr{m_2}{v_1}a_1P_R+ \fr{m_a}{v_3}a_3P_L\right)\fr{i(/\!\!\!p_1+m_2)}{p_1^2-m_2^2}\left(\fr{im_2}{v_1}\fr{c_\al}{\sqrt{2}}\right) v_2\times\fr{i}{D_1}\crn
&=&\sum_{a}\left[-\fr{\sqrt{2}c_\al}{v_1}\right] \fr{m_2}{m_1^2-m_2^2}V_{1a}V^*_{2a}  \crn
&\times&\int\fr{d^4k}{(2\pi)^4}\times\left[\fr{m_1m_2}{v^2_1}a^2_1\fr{\bar{u}_1/\!\!\!k/\!\!\!p_1P_Lv_2}{D_0D_1}+\fr{m_1m^2_2}{v^2_1}a^2_1\fr{\bar{u}_1/\!\!\!k P_Rv_2}{D_0D_1}\right.\crn
&& \left. +\fr{m_1m^2_a}{v_1v_3}a_1a_3\fr{\bar{u}_1/\!\!\!p_1P_Rv_2}{D_0D_1}+\fr{m_1m_2m^2_a}{v_1v_3}a_1a_3 \fr{\bar{u}_1P_Lv_2}{D_0D_1}+\fr{m^2_a}{v^2_3}a^2_3\fr{\bar{u}_1/\!\!\!k/\!\!\!p_1P_Rv_2}{D_0D_1}\right.\crn
&&\left. +\fr{m_2m^2_a}{v^2_3}a^2_3\fr{\bar{u}_1/\!\!\!k P_Lv_2}{D_0D_1}+ \fr{m_2m^2_a}{v_1v_3}a_1a_3\fr{\bar{u}_1/\!\!\!p_1 P_Lv_2}{D_0D_1}+\fr{m^2_2m^2_a}{v_1v_3}a_1a_3\fr{\bar{u}_1 P_Rv_2}{D_0D_1} \right]\crn
&=&\sum_{a}\left[-\fr{\sqrt{2}c_\al}{v_1}\right] \fr{m_2}{m_1^2-m_2^2}V_{1a}V^*_{2a}
\crn &&\times \left\{\fr{}{}\bar{u}_1P_Lv_2\times m_1m_2\left[ \fb{$\bf 2m_a^2\fr{a_1a_3}{v_1v_3}B^{(1)}_0+m_a^2\fr{a_3^2}{v_3^2}B^{(1)}_1$} +\fr{m_1^2}{v_1^2}a_1^2B^{(1)}_1\right]\right.\crn
&&\left.+\bar{u}_1P_Rv_2\left[\fr{}{}\fb{$\bf m_a^2\fr{a_1a_3}{v_1v_3}(m_1^2+m_2^2)B^{(1)}_0  + m_1^2m_a^2\fr{a_3^2}{v_3^2}B^{(1)}_1$} +\fr{m_1^2m_2^2}{v_1^2}a_1^2B^{(1)}_1\fr{}{}\right]\right\}.\crn \label{Di1i}\eea}
The contribution from the diagram \ref{nDiagram}k) is:
{\footnotesize\bea
i\mathcal{M}^{HF}_{(k)}&=&\sum_{a}\int\fr{d^4k}{(2\pi)^4} \times \bar{u}_1\left(\fr{im_1c_\al}{v_1\sqrt{2}}\right) \crn
&\times&\fr{i(-/\!\!\!p_2+m_1)}{p_2^2-m_1^2}(-i\sqrt{2}V_{1a})\left(\fr{m_1}{v_1}a_1P_L+ \fr{m_a}{v_3}a_3P_R\right)\crn
&&\times\fr{i(/\!\!\!k+m_a)}{D_0 }(-i\sqrt{2}V^*_{2a})\left(\fr{m_2}{v_1}a_1P_R+ \fr{m_a}{v_3}a_3P_L\right)v_2\times\fr{i}{D_2}\crn
&=&\sum_{a}\left(-\fr{i\sqrt{2}c_\al}{v_1}\right)\fr{m_1}{m_2^2-m_1^2}V_{1a}V^*_{2a}\crn
&\times& \int\fr{d^4k}{(2\pi)^4} \times\left[-\fr{m_1m_2}{v^2_1}a^2_1\fr{\bar{u}_1/\!\!\!p_2/\!\!\!kP_Rv_2}{D_0D_2} +\fr{m^2_1m_2}{v^2_1}a^2_1\fr{\bar{u}_1/\!\!\!k P_Rv_2}{D_0D_2}\right.\crn
&& \left.- \fr{m_1m_a^2}{v_1v_3}a_1a_3\fr{\bar{u}_1/\!\!\!p_2P_Lv_2}{D_0D_2}+\fr{m^2_1m^2_a}{v_1v_3} a_1a_3\fr{\bar{u}_1P_Lv_2}{D_0D_2}-\fr{m^2_a}{v^2_3}a^2_3\fr{\bar{u}_1/\!\!\!p_2/\!\!\!k P_Lv_2}{D_0D_2}\right.\crn
&&+\left. \fr{m_1m^2_a}{v^2_3}a^2_3\fr{\bar{u}_1/\!\!\!k P_Lv_2}{D_0D_2}- \fr{m_2m^2_a}{v_1v_3}a_1a_3\fr{\bar{u}_1/\!\!\!p_2 P_Rv_2}{D_0D_1}+\fr{m_1m_2m^2_a}{v_1v_3}a_1a_3\fr{\bar{u}_1 P_Rv_2}{D_0D_1} \right]\crn
&=&\sum_{a}\left(-\fr{i\sqrt{2}c_\al}{v_1}\right)\fr{m_1}{m_2^2-m_1^2}V_{1a}V^*_{2a} \left\{\fr{}{} \bar{u}_1P_Lv_2\left[\fr{}{}\fb{$\bf\fr{m_a^2}{v_1v_3}a_1a_3(m_1^2+m_2^2)B^{(2)}_0$}\right.\right.\crn
&&\left.\left. -\fb{$\bf\fr{m_2^2m_a^2}{v_3^2}a_3^2 B_1^{(2)}$}- \fr{m_1^2m_2^2}{v_1^2}a_1^2B^{(2)}_1 \right]\right.\crn
&&\left. +\bar{u}_1P_Rv_2\times m_1m_2\left[\fr{}{}\fb{$\bf 2\fr{m_a^2}{v_1v_3}a_1a_3 B_0^{(2)}-\fr{m_a^2}{v_3^2}a_3^2B^{(2)}_1$} -\fr{m_2^2}{v_1^2}a_1^2B^{(2)}_1\fr{}{}\right]\right\}.\crn \label{Di1k}
\eea}
The total amplitude from the two diagrams \ref{nDiagram}i) and k) is:
{\footnotesize\bea
i\mathcal{M}^{FH}_{(i+k)}&=&\sum_{a}V_{1a}V^*_{2a} \left[-\fr{i\sqrt{2}c_\al}{v_1}\right] \times\left\{\bar{u}_1P_Lv_2\times \fr{m_1}{m_1^2-m_2^2}\right.\crn
&&\left.\times \left[m^2_1m^2_2
\fr{a_1^2}{v_1^2}\left(B_1^{(1)}+B_1^{(2)}\right) +\fb{$\bf m^2_a\fr{a_1a_3}{v_1v_3}\left(2m^2_2B_0^{(1)}-(m^2_1 +m^2_2)B_0^{(2)}\right)$}\right.\right.\crn
&&\left.\left. +\fb{$\bf m^2_2m^2_a\fr{a^2_3}{v^2_3}\left(B_1^{(1)}+B_1^{(2)}\right)$}\fr{}{}\right]+\bar{u}_1P_Rv_2\times \fr{m_2}{m_1^2-m_2^2}\right.\crn
&&\times\left. \left[m^2_1m^2_2\fr{a_1^2}{v_1^2}\left(B_1^{(1)}+B_1^{(2)}\right)+ \fb{$\bf m^2_1m^2_a\fr{a^2_3}{v^2_3}\left(B_1^{(1)}+B_1^{(2)}\right)$}\right.\right.\crn
&&\left.\left. +\fb{$\bf m^2_a\fr{a_1a_3}{v_1v_3}\left(-2m^2_1B_0^{(2)}+(m^2_1 +m^2_2)B_0^{(1)}\right)$}\fr{}{}\right]\right\}.\crn
\eea}
The final result is written as
\be  \mathcal{M}^{FH}_{(ik)} =\left[-\fr{8c_\al}{64\pi^2\sqrt{2}}\right] \sum_{a}V_{1a}V^*_{2a} \left[ \left(\overline{u_1} P_L v_2\right)   E^{FH}_L +  \left( \overline{u_1} P_R v_2\right)  E^{FH}_R \right],  \label{Diep} \ee
where $E^{FH}_{L,R}$ are defined in (\ref{DfhL}) and (\ref{DfhR}).
After calculating contributions from  all diagrams with virtual neutral leptons $N_a$ we can prove that   all divergent parts containing
 the factor $m_a^2$ will be canceled  in the total contribution.  The details are shown below. For active neutrinos the calculation is the same.
\subsection{Particular calculation for canceling divergence}
In this section, for contribution of  exotic neutral leptons $N_a$ we use the following relations
\bea
a_1 &\rightarrow&  c_\theta,\;  a_2\rightarrow a_3 = s_\theta, \; v_1 = \fr{2m_V}{g}s_\theta, \;  v_3 = \fr{2m_V}{g}c_\theta, \crn
\fr{a_1}{v_1} &=& \fr{g}{2m_V}\fr{c_\theta}{s_\theta}, \; \fr{a_3}{v_3}=\fr{g}{2m_V}\fr{s_\theta}{c_\theta}, \; \fr{a_1a_3}{v_1v_3}=\fr{g^2}{4m^2_V}. \label{aijNlepton1}
\eea
And we concentrate on the divergent parts which are bolded in the expressions of  the amplitudes calculated above. With the notations of the divergences shown in the appendix \ref{mainte}, all of  divergent parts are collected as follows,
 \bea \mathrm{Div}\left[\mathcal{M}_{(a)}^{FVV}\right] &=& B\times \left[c_{\alpha}\times (-3s_{\theta})+\sqrt{2}s_{\alpha}(3c_{\theta}) \right], \crn   \mathrm{Div}\left[\mathcal{M}_{(b+c)}^{FHV}\right]&=&  B\times \left[ c_{\alpha}\times\frac{s^2_{\theta}-2c^2_{\theta}}{s_{\theta}} + \sqrt{2}s_{\alpha}\times\frac{s^2_{\theta}-2c^2_{\theta}}{c_{\theta}}\right], \crn
\mathrm{Div}\left[\mathcal{M}_{(e)}^{VFF}\right]&=& B\times \sqrt{2}s_{\alpha}\times\frac{-3}{c_{\theta}},\crn
\mathrm{Div}\left[\mathcal{M}_{(f)}^{HFF}\right]&=& B\times  \sqrt{2}s_{\alpha}\times\frac{2}{c_{\theta}} ,\crn
\mathrm{Div}\left[\mathcal{M}^{FV}_{(g)} \right]&=&\fr{1}{m_1^2-m_2^2}\left[m_2^2B_L+m_1^2B_R\right]\times \fr{3c_\al}{s_\theta},\crn
\mathrm{Div}\left[\mathcal{M}^{FV}_{(h)} \right]&=&\fr{1}{m_1^2-m_2^2}\left[m_2^2B_L+m_1^2B_R\right]\times \fr{-3c_\al}{s_\theta},\crn
\mathrm{Div}\left[\mathcal{M}_{(i+k)}^{FH}\right]&=& B\times  c_{\alpha}\times\frac{2}{s_{\theta}}, \label{divergentp}\eea
where
\bea
B&=&\fr{g^3}{128\pi^2}\fr{m^2_{\nu_{a}}}{m_W^3}\times \Delta_\epsilon\times \left[ \bar{u}_1P_Lv_2\times m_1+\bar{u}_1P_Rv_2\times m_2\right]\crn
B_L&=&\fr{g^3}{128\pi^2}\fr{m^2_{\nu_{a}}}{m_W^3}\times \Delta_\epsilon\times\bar{u}_1P_Lv_2\times m_1, \;
B_R= \fr{g^3}{128\pi^2}\fr{m^2_{\nu_{a}}}{m_W^3}\times \Delta_\epsilon\times\bar{u}_1P_Rv_2\times m_2.\nn
\eea
It is easy to see that the sum over all factors is zero. Furthermore,  it is interesting to see that the sums of the two parts having factor $c_{\alpha}$ and $\sqrt{2}s_{\alpha}$  independently result the zero values. From (\ref{ecpeh0}),  the factor $c_{\alpha}$  arises from the contributions of  neutral components of $ \eta$ and $ \rho$, while the $s_{\alpha}$   factor arises from the contribution of $\chi$.

 For contribution of the active  neutrinos,  the two diagrams  $(b)$ and $(c)$ of  the fig.\ref{nDiagram} do not give contributions due to absence of the $H^-_2H^+_2W$
 couplings. Using the following properties
\be
a_1 = 1, a_2 = 1, v_1 =v_2= \fr{2m_W}{\sqrt{2}g} , \; \fr{a_1}{v_1}=\fr{a_2}{v_2}=\fr{\sqrt{2}g}{2m_W},\; \fr{a_1a_2}{v_1v_2}=\fr{g^2}{2m^2_W},\non
\ee
 we list the non-zero divergent terms of  the relevant diagrams as follows
\bea
\mathrm{Div}\left[\mathcal{M}^{FVV}_{(a)} \right]&=&\mathcal{B}\times (-3c_\al),\crn
\mathrm{Div}\left[\mathcal{M}^{VFF}_{(e)} \right]&=&\mathcal{B}\times (3c_\al),\crn
\mathrm{Div}\left[\mathcal{M}^{HFF}_{(f)} \right]&=&\mathcal{B}\times (-2c_\al),\crn
\mathrm{Div}\left[\mathcal{M}^{FV}_{(g)} \right]&=&\fr{1}{m_1^2-m_2^2}\left[m_2^2\mathcal{B'}_L+m_1^2\mathcal{B'}_R\right]\times (c_\al),\crn
\mathrm{Div}\left[\mathcal{M}^{FV}_{(h)} \right]&=&\fr{1}{m_1^2-m_2^2}\left[m_2^2\mathcal{B'}_L+m_1^2\mathcal{B'}_R\right]\times (-c_\al),\crn
\mathrm{Div}\left[\mathcal{M}^{FH}_{(i)} \right]&=&\fr{-c_\al}{m_1^2-m_2^2}\left[5m_2^2\mathcal{B'}_L+(3m_1^2+2m_2^2)\mathcal{B'}_R\right],\crn
\mathrm{Div}\left[\mathcal{M}^{FH}_{(k)} \right]&=&\fr{c_\al}{m_1^2-m_2^2}\left[(2m_1^2+3m_2^2)\mathcal{B'}_L+5m_1^2\mathcal{B'}_R\right],\crn
\mathrm{Div}\left[\mathcal{M}^{FH}_{(i+k)} \right]&=&\mathcal{B}\times (2c_\al), \nonumber \eea
where
\bea
\mathcal{B}&=&\fr{g^3}{128\pi^2}\fr{m^2_{\nu_{a}}}{m_W^3}\times \Delta_\epsilon\times \left[ \bar{u}_1P_Lv_2\times m_1+\bar{u}_1P_Rv_2\times m_2\right]\crn
\mathcal{B'}_L&=&\fr{g^3}{128\pi^2}\fr{m^2_{\nu_{a}}}{m_W^3}\times \Delta_\epsilon\times\bar{u}_1P_Lv_2\times m_1, \;
\mathcal{B'}_R =\fr{g^3}{128\pi^2}\fr{m^2_{\nu_{a}}}{m_W^3}\times \Delta_\epsilon\times\bar{u}_1P_Rv_2\times m_2.\nonumber
\eea
We see again that sum of all divergent terms is zero.

\end{document}